\documentclass[onecolumn]{emulateapj}

\usepackage{amsmath, amssymb}
\usepackage{graphicx,color,colordvi} 
\usepackage{latexsym} 
\usepackage{url} 
\usepackage{times} 

\def \bmath #1 {{\hbox{\boldmath{$#1$}\unboldmath}}} 

\slugcomment{Submitted to Astrophys. J.}

\shorttitle{Depolarizing collisions}

\shortauthors{Manso Sainz et al.}

\begin{document}

\title{Depolarizing Collisions with Hydrogen:  
Neutral and Singly Ionized Alkaline Earths}

\author{ Rafael Manso Sainz$^a$, Octavio Roncero$^{b,1}$, Cristina Sanz-Sanz$^{b,c}$, Alfredo Aguado$^c$, 
  Andr\'es Asensio Ramos$^a$, and Javier Trujillo Bueno$^a$}\footnote{e-mail: octavio.roncero@csic.es}

\affil{$^a$ Instituto de Astrof\' isica de Canarias, V\' ia L\'actea s/n, E-38205 La Laguna, Tenerife, Spain}                 
\affil{Departamento de Astrof\'\i sica, Universidad de La Laguna, E-38206, Tenerife, Spain}                              

\affil{$^b$  Instituto de F{\'\i}sica Fundamental 
(IFF-CSIC), C.S.I.C., Serrano 123, 28006 Madrid, Spain}

\affil{$^c$ Departamento de Qu{\'\i}mica F{\'\i}sica, Unidad Asociada UAM-CSIC, Facultad de Ciencias M-14, 
Universidad Aut\'onoma de Madrid, 28049, Madrid, Spain}

\begin{abstract}
Depolarizing collisions are elastic or quasielastic collisions that equalize the 
populations and destroy the coherence between the magnetic sublevels of atomic levels.
In astrophysical plasmas, the main depolarizing collider is neutral hydrogen.
We consider depolarizing rates on the lowest levels of neutral and singly ionized 
alkaly-earths  
Mg~{\sc i}, Sr~{\sc i}, Ba~{\sc i}, Mg~{\sc ii}, Ca~{\sc ii}, and Ba~{\sc ii},
due to collisions with H$^\circ$.
We compute {\em ab initio} potential curves of the atom-H$^\circ$ system 
and solve the quantum mechanical dynamics.
From the scattering amplitudes we calculate the depolarizing rates for Maxwellian
distributions of colliders at temperatures $T\leq $10000~K.
A comparative analysis of our results and previous calculations in the literature is done.
We discuss the effect of these rates on the formation of scattering 
polarization patterns of resonant lines of alkali-earths in the solar atmosphere,
and their effect on Hanle effect diagnostics of solar magnetic fields.
\end{abstract}

\keywords{stars: magnetic fields --- techniques: polarization --- 
methods: data analysis, statistical}

\maketitle 
 
\section{Introduction} 
 
Since the discovery of the linearly polarized component of the Fraunhofer spectrum 
observed close to the solar limb \citep{Stenflo+83a, Stenflo+83b},
much effort has been devoted to understand the physical mechanisms involved in its 
formation, which is dominated by scattering in the continuum and spectral lines
 \citep[][]{Stenflo94, Stenflo97, TrujilloLandi97, FluriStenflo99, Trujillo+02b, Innocenti:04}.
Special attention has received, observationally and theoretically, 
the influence of weak or tangled magnetic fields on resonance line polarization
through the Hanle effect  
\cite[][]{MoruzziStrumia91},
which has opened a new diagnostic window for the magnetism in the solar atmosphere
\cite[e.g., ][]{Stenflo82, Stenflo91, Leroy89, Bommier+94, Faurobert93, Faurobert+95, 
Lin+98, Trujillo01, TrujilloManso02, LopezAristeCasini05, Trujillo+05}.
By contrast, elastic depolarizing collisions have received relatively little
attention in this context, a neglect motivated by their apparent lack of diagnostic value.
However, collisions compete with magnetic fields to depolarize the atomic levels, 
which must be properly accounted for to calibrate Hanle effect diagnostic techniques;
besides, by broadening the atomic energy levels, they modulate the magnetic field 
strength at which the Hanle effect is sensitive \citep{Lamb70}.

Alkaline-earth metals, atomic and singly ionized, show 
some of the strongest resonant lines in the Fraunhofer spectrum and
their resonance polarization patterns have proved remarkable too.
Their interpretation has posed important theoretical challenges. 
Thus, the Ca~{\sc ii} infrared triplet  
and the Mg~{\sc i} $b$-lines \citep[][]{Stenflo+00}
provided the first clear manifestation of the presence of atomic polarization in metastable 
levels in the solar chromosphere \citep[][{ consistently with our results here, see Section 4}]{MansoTrujillo03, Trujillo01};
the polarization pattern around the $H$- and $K$-lines of Ca~{\sc ii} 
 \citep{Stenflo+80, Stenflo80}, and the $h$ and $k$-lines of Mg~{\sc ii} 
 \citep{HenzeStenflo87}, arise from the interference between the upper ${}^2P$ levels
 \citep{Stenflo80, BelluzziTrujillo12}.
{ Resonance lines of alkali-earths have proved essential to diagnose} unresolved fields in the solar atmosphere.
The remarkable scattering polarization signal in the Sr~{\sc i} at 460.7~nm 
 \citep{Stenflo+80, Faurobert+01, BommierMolodij02}, in particular,
has been extensively observed with the aim of diagnosing tangled 
magnetic fields in the solar atmosphere
 \citep{Stenflo82, Faurobert93, Faurobert94, Faurobert+95, Bianda+99, Trujillo+04}.

Depolarizing collisions in alkali and alkaline earth atoms with a foreign 
noble gas have been throughly studied theoretically and experimentally 
 \citep[see][and references therein]{LambterHaar71, Omont77, Baylis78}.
However, in the solar atmosphere the most important depolarizing collider 
is neutral hydrogen \citep{Lamb70}.
Rough estimates for the depolarizing rates can be obtained from
a dipole-dipole van der Waals approximation for the atom-H$^\circ$
 \citep[][see also Landi Degl'Innocenti \& Landolfi 2004]{LambterHaar71}.
More recently, \cite{Derouich+03, Derouich+03b, 
Derouich+04, Derouich+04b, Derouich+05, DerouichBarklem07} 
used {\em ab initio} interaction potentials and a semi-classical
approach using straight line trajectories at different impact parameters.
\cite{Kerkeni+00, Kerkeni02, KerkeniBommier02, Kerkeni+03}
have calculated depolarizing rates with neutral hydrogen
with a fully quantum mechanical approach for the interaction
Hamiltonian and dynamics.
We follow a similar approach here.
We computed {\em ab initio} potential curves of the atom-H$^\circ$ system
and then solved the dynamics within the scattering matrix formalism.
From them, we calculated the depolarizing collision rates for
a Maxwellian distribution of colliders.

\begin{figure}[t]
\centering\includegraphics[scale=0.8]{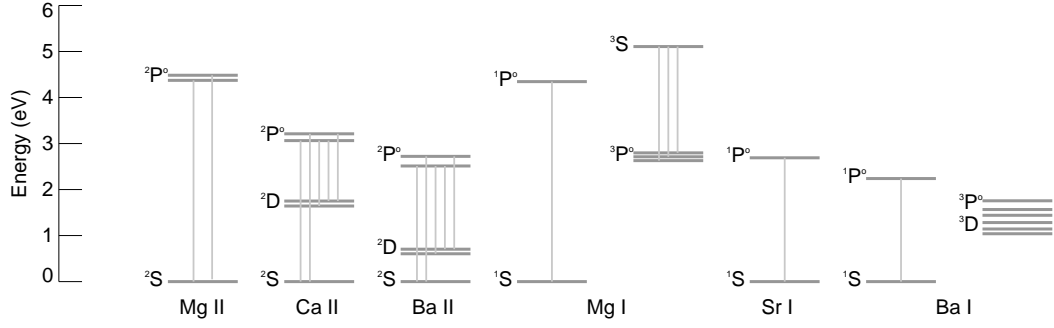} 
\caption{\label{solar-transitions}
Partial Grotrian diagrams with the lowest energy levels of 
the most abundant alkaline earth metals. 
All levels (except the ${}^2S$ and ${}^2D$ terms of the Mg~{\sc i}
at 8.6 and 8.8 ev) and resonance transitions between them considered in this work
are represented.
 The fine structure energy splitting in Mg and Ca are exaggerated for
clarity.}
\vspace*{0.5cm}
\end{figure}

The most important results of the present work are summarized in Table~\ref{depolarization-constants}, 
which gives the depolarizing rates for the lowest-lying energy levels of
Mg~{\sc i}, Sr~{\sc i}, Ba~{\sc i}, Mg~{\sc ii}, Ca~{\sc ii}, and Ba~{\sc ii}.
Figure~\ref{sunny} shows the relative importance of these depolarizing rates 
in the solar atmosphere as compared to the radiative (polarizing) rates
in the main resonance lines of these atoms and ions (Figure~\ref{solar-transitions}). 

The next section introduces the general theoretical framework used  
and Section~3 details the calculations performed for each individual atom and ion considered.
The impact on the formation of the scattering polarization patterns in the solar
atmosphere is discussed in Section~4.

%

\section{Theoretical methodology for Collisional rates} 

%

{ A complete treatment of the Hanle effect on an atom A requires the resolution of the
master equations considering radiative transitions, collisional transitions, and the interaction with the magnetic fields. 
Such formalism has been presented by several authors \citep[e.g.,][]{Bommier-Sahal-Brechot:78,Bommier:80,Innocenti:04}.
 In this work we focus only on deriving the collisional rates caused by collisions with neutral hydrogen atoms,
 which may have a significant depolarization effect. These collisions are of the type}
 
\begin{eqnarray}
A(\alpha,J,M)+ H(S,M_S) \rightarrow A(\alpha',J',M') +  H(S,M'_S),
\end{eqnarray}
between an atom $A$ in a state $\left\vert \alpha,  J,M\right\rangle$
with total angular momentum $J$, magnetic quantum number $M$, and energy $E^A_{\alpha,J}$
($\alpha$ comprises all aditional quantum numbers to characterize the state), 
and a hydrogen atom $H$ in its ground state, with total angular momentum $S=1/2$,
and magnetic quantum number $M_S$.
{\bf We consider only low energy collisions ($E\ll 10$~eV), for which the 
perturber atom (neutral hydrogen) 
remains on the ground level after the collision, and 
the state of the atom under consideration 
may change only between levels of the same $LS$ term (i.e., $\alpha=\alpha'$).
We call {\em elastic} collisions those for which $J'=J$, and {\em quasi-elastic}
collisions those for which $J'\ne J$. }

The density matrix  of the two atoms system is represented as
$$
\left\langle \alpha_1 J_1 M_1;  S_1 M_S^1; {\bf k}_1 \left\vert \hat \rho \right\vert
 \alpha_2 J_2 M_2;   S_2 M_S^2; {\bf k}_2 \right\rangle,
$$
where ${\bf k}=\hbar{\bf v} $ is the wavevector, parallel to the relative velocity vector, 
${\bf v}$, between the A and H atoms,
of modulus  $k=\sqrt{2\mu (E-E^A_{\alpha,J}-E^H_{S})/\hbar^2}$,
with $\mu= m_A m_H/(m_A+m_H)$ being
the reduced mass of the colliding atoms and $E$ the total energy of the system.
Our interest here focus on the polarization of the atom A, 
for which we consider its coherence between
$J_1, M_1$ and $J_2, M_2$ states. 
Neutral hydrogen atoms, with density  $N_{H^\circ}$, interact  isotropically with 
the $A$ atoms, and all the $M_S$ sublevels have the weight $1/(2S+1)$. 
{\bf Since we are not interested on the final state of the hydrogen atoms, 
we sum over all the final states of H$^\circ$ by simply making $S_1=S_2$
and $M_S^1=M_S^2$ (i.e., we consider only diagonal terms in the density matrix).}
Under these conditions, the reduced density matrix for the $A$ subsystem can be written as 
\begin{eqnarray}\label{density-matrix}
\rho_{\alpha J} (M_1, M_2)
=\sum_{M_S} {1\over 2S+1} \int v^2 dv f(T,v) \int d{\hat k}  
\,\left\langle \alpha J M_1;  \beta S M_S; {\bf k}
 \left\vert \hat \rho \right\vert
 \alpha J M_2;  \beta S M_S; {\bf k} \right\rangle,
\end{eqnarray}
where $f(T,v)$ is {\bf the velocity distribution of the perturbers,} 
here the usual Maxwell-Boltzmann one,
${\hat k}\equiv {\hat v}$ denotes the polar and azimuthal angles of ${\bf k}$ vector
with respect to the frame of the observation. In the above equation the coherence terms
between states of A atom with different $\alpha J$ states are neglected. 
The multipole moments of the density matrix are then defined as 
\begin{eqnarray}\label{density-matrix-multipoles}
\rho^K_Q(\alpha J) = \sum_{M_1 M_2} (-1)^{J-M}\sqrt{2K+1} 
      \left(\begin{array}{ccc} 
             J & J & K\\ 
             M_1 &-M_2 & -Q 
      \end{array}\right)
\rho_{\alpha J} (M_1,M_2), 
\end{eqnarray}
where the notation used by Degl'Innocenti \citep{Innocenti:04} has been adopted.

{ The contribution of collisional transitions to the rate of change of the density matrix elements 
can be written as} \citep{Bommier:80,Innocenti:04}

\begin{eqnarray}\label{master-equation-densities}
&& \frac{d}{dt} \rho_{\alpha J} (M_1, M_2)
= \sum_{\alpha' J' M'_1 M'_2} 
  C(\alpha J M_1 M_2\leftarrow \alpha' J' M'_1 M'_2 )  \quad \rho_{\alpha' J'} (M'_1, M'_2) \nonumber\\
&& -{1\over 2}\sum_{\alpha' J' M' M_3} \Big\lbrack  
 C( \alpha' J' M' M'\leftarrow \alpha J M_3 M_1)\quad \rho_{\alpha J} (M_3, M_2) 
+ C(\alpha' J' M' M'\leftarrow \alpha J M_2 M_3 ) \quad\rho_{\alpha J} (M_1, M_3) 
\Big\rbrack, 
\end{eqnarray}
where $C(\alpha' J' M'_1 M'_2\leftarrow \alpha J M_1 M_2 )$ are the corresponding rate constants
at a given temperature,  defined as
\begin{eqnarray}\label{rates}
C( \alpha' J' M'_1 M'_2\leftarrow \alpha J M_1 M_2) = 
N_{H^\circ}   \int v^2 dv f(T,v)\,
\sigma^0(\alpha J M_1 M_2; \alpha' J' M'_1 M'_2).
\end{eqnarray}
In this equation $\sigma^0$ is the collisional cross-section averaged over the isotropic distribution
of H atoms depending on the collisional energy defined 
as \citep{Balint-Kurti-Vasyutinskii:09,Gonzalez-Sanchez-etal:11,Krasilnikov-etal:13}
\begin{eqnarray}\label{isotropic-cross-section}
\sigma^0(\alpha J M_1 M_2; \alpha' J' M'_1 M'_2) = 
\sum_{M _S, M'_S} {1\over 2S+1} \int d{\hat k}  \int d{\hat k}'  
\quad  [f^{\alpha J M_1 S M_S }_{\alpha' J' M'_1 S' M'_S}({\hat k},{\hat k}')]^*
\quad f^{\alpha J M_2 S M_S}_{\alpha' J' M'_2 S' M'_S}({\hat k},{\hat k}'),
\end{eqnarray}
with  $f$ being the scattering amplitude, defined below, which relates
the amplitude of the products in state $(\alpha' J' M'_1 S' M'_S)$
 scattered on a final direction ${\hat k}'$ from 
an initial state  $(\alpha J M_1 S M_S)$ colliding with a relative speed
parallel to ${\hat k}$.

Equations~(\ref{master-equation-densities}) describing the evolution of the density matrix of an
atom under the only action of collisions can be simply added to 
the equations describing the evolution under radiative processes
and external fields \citep[as given, for example, by ][]{Bommier:80,
Bommier-Sahal-Brechot:78, Innocenti:04},
if the impact approximation is valid. 
This requires that the collision time be much smaller than
the relaxation time due to radiative rates \citep[see ][]{LambterHaar71},
which is satisfied in most astrophysical and laboratory plasmas.

The collisional rates can be written as
 \citep{Omont77,Innocenti:04}
\begin{equation}\label{rate-multipoles}
C( \alpha' J' M'_1 M'_2\leftarrow \alpha J M_1 M_2)=(-1)^{J-M_2+J'-M'_2}\sqrt{{2J+1 \over 2J'+1}}
\sum_{K}(2K+1)
\left(\begin{array}{ccc} 
             J' & J' & K\\ 
             M'_2 &-M'_1 & Q
      \end{array}\right)
      \left(\begin{array}{ccc} 
             J & J & K\\ 
             M_2 &-M_1 & Q 
      \end{array}\right)
{C}^{(K)}(\alpha' J'\leftarrow \alpha J )
\end{equation}
{\bf where ${C}^K( \alpha' J'\leftarrow \alpha J)$ are the multipole components of the collisional rates, which are independent of $Q$ because of the assumed isotropic conditions of the collisional processes \citep{Omont77}. Note that the expression of Eq.~(\ref{rate-multipoles}) 
differs from that given by \citet{Omont77} in that we have introduced the factor  
$\sqrt{{2J+1 \over 2J'+1}}$ to follow the same collisional rate definitions as in Section 7.13 of \cite{Innocenti:04}. The only differences with respect to the notation used by \cite{Innocenti:04} are that we do not make any notational distinction between downwards and upwards collisional transitions and our use of arrows to indicate the initial and final states. Conversely, 

\begin{equation}\label{rate-multipoles-inverse}
C^{(K)}( \alpha' J'\leftarrow \alpha J )= \sqrt{\frac{2J'+1}{2J+1}}
\sum_{M_1 M_2 M'_1 M'_2 P} (-1)^{J'-M'_2+J-M_2}
\left(\begin{array}{ccc} 
             J' & J' & K\\ 
             M'_2 & -M'_1 & P 
      \end{array}\right)
      \left(\begin{array}{ccc} 
             J & J & K\\ 
             M_2 & -M_1 & P 
      \end{array}\right)
 C( \alpha' J' M'_1 M'_2\leftarrow \alpha J M_1 M_2).
\end{equation}

The master equations in terms of the multipole components is given 
by Eq.(7.101) of \cite{Innocenti:04}, which reads
\begin{equation}\label{master-equation-densities-K}
\frac{d}{dt}\rho^K_Q(\alpha J) =
\sum_{J' \neq J} \sqrt{\frac{2J'+1}{2J+1}}C^{(K)}(\alpha J\leftarrow \alpha' J')\rho^K_Q(\alpha' J')
-\left[ \sum_{\alpha' J'\neq \alpha J} C^{(0)}(\alpha' J'\leftarrow \alpha J)
+ D^{(K)}(\alpha J)\right] \rho^K_Q(\alpha J),
\end{equation}
where
\begin{equation}\label{elastic-depolarization-rates}
D^{(K)}(\alpha J)=C^{(0)}(\alpha J\leftarrow  \alpha J)-C^{(K)}(\alpha J\leftarrow  \alpha J)
\end{equation}
are the elastic depolarization rates, as defined in equation (7.102) of \cite{Innocenti:04}. 

The total depolarizing rate of the $\alpha J$ level 
is defined as the term within the brackets in Eq.~(\ref{master-equation-densities-K}):
\begin{equation}\label{depolarizing-rate}
g^K(\alpha J)= \bar{C}(\alpha J)+D^K(\alpha J),
\end{equation}
where
\begin{equation}\label{inelastic-rate-moments}
\bar{C}(\alpha J)= \sum_{\alpha' J'}
C^{(0)}( \alpha' J'\leftarrow \alpha J). 
\end{equation}
are the inelastic rates.
}

\subsection{Scattering amplitudes and cross sections}

The scattering wave functions for inelastic 
collisions for incoming plane waves is subject to the boundary
conditions \citep{Curtiss-Adler:52}
\begin{equation}
\begin{split}
\lim_{R\rightarrow\infty} \Phi_{\alpha J M S M_S}({\boldsymbol k},{\hat k}') =& 
e^{i{\boldsymbol k} \cdot {\boldsymbol R}}\, \vert \alpha J M \rangle  \vert S M_S\rangle  
\\ 
&+ i \sum_{\alpha'J' M'  S'  M'_S}    f^{\alpha J M S M_S *}_{\alpha' J' M' S' M'_S}({\hat k},{\hat k}')
{e^{i k' R}\over R \sqrt{k' k_\alpha}}\,  \vert \alpha' J' M' \rangle  \vert S' M'_S\rangle
\end{split}
\end{equation}
which determines the scattering amplitude. This $f$ corresponds to a single 
collision and therefore does not present isotropic symmetry but cylindrical
around the incoming velocity ${\hat k}$. 

The scattering amplitude  depends on the three coordinates
of ${\bf k}$, involving the resolution of a set of three-dimensional differential
equations. To reduce the problem, it is convenient to perform a partial wave expansion, $i.e.$
expand the wave function in terms of eigenfunctions of the total angular momentum, ${\bf J}_t=
\bmath {\ell} + {\bf j}$, where $\bmath {\ell} $ is the orbital angular momentum of H with respect to
A and ${\bf j} = {\bf J} + {\bf S }$ is the angular momentum of the two fragments. 
Using the partial wave expansion of the incident plane wave, 
 after some  
algebra \citep{Arthurs-Dalgarno:60,Rowe-McCaffery:79,Alexander-Davis:83b,Child-book}
the scattering amplitude become
\begin{equation}\label{scattering-amplitude} 
\begin{split}
 f^{\alpha J M S M_S}_{\alpha' J' M S' M'_S}({\hat k},{\hat k}') =& 
\sum_{J_t M_t}\sum_{jj'}\sum_{mm'}\sum_{\ell m_\ell}\sum_{\ell' m'_\ell} {2\pi\over k} (-1)^\ell  
            \,  Y^*_{\ell m_\ell}({\hat k})\,Y_{\ell' m'_\ell}({\hat k}')
          \,T^{J_t}_{n,n'}(E) 
\\ 
&\times  (-1)^{j+j'+J+J'-\ell-\ell'-2S+2M_t+m+m'}(2J_t+1)\sqrt{(2j+1)(2j'+1)} \\
&\times\left(\begin{array}{ccc}
j & \ell & J_t \\ 
m & m_\ell   & -M_t 
\end{array}\right) 
\left(\begin{array}{ccc}
j' & \ell' & J_t \\
 m' & m'_\ell  & -M_t 
\end{array}\right) 
\left(\begin{array}{ccc}
J & S & j \\ 
M & M_S   & -m
\end{array}\right) 
\left(\begin{array}{ccc}
J' & S & j' \\ 
M' & M'_S   & -m' 
\end{array}\right),
\end{split}
\end{equation} 
where we have introduced the collective quantum number $n\equiv j \ell \alpha J S$ to simplify
the notation, and 
where $M_t$ and $m$ are the projections of ${\bf J}_t$ and ${\bf j}$
on the z-axis. In this expression $T^{J_t}_{n, n'}(E) =  
\delta_{nn'} -S^{J_t}_{n,n'}(E)$,
with $S^{J_t}_{n,n'}(E)$ being the scattering S-matrix, which provides all
the information about the collision event for a given total angular momentum $J_t$. These S-matrices
are obtained by integrating a set of coupled differential equations, as described below.

Following \cite{Alexander-Davis:83b}, it is convenient to expand the 
scattering amplitude in terms of spherical tensors, as for the density matrix,
 whose  state-multipoles are given by 
\begin{equation}\label{scattering-amplitud-state-multipoles} 
\begin{split}
f^{K''}_{Q''} (\alpha J S M_S, {\hat k}\rightarrow\alpha' J' S' M'_S, {\hat k}') =& 
 \sum_{\ell m_\ell} \sum_{\ell' m'_\ell} \sum_{K Q} \sum_{K' Q'}  b^{\ell K}_{\ell' K'}(K'')
  Y_{\ell m_\ell}({\hat k})\,Y_{\ell' m'_\ell}({\hat k}') 
\\ 
&\times 
(-1)^{\ell+2K''+2M_S+2M'_S-Q+2Q''} \,(2K+1)(2K'+1)\sqrt{2K''+1} 
\\
&\times
\left(\begin{array}{ccc} 
   K' & K & K''\\ 
   Q' & -Q & -Q'' 
\end{array}\right) 
\left(\begin{array}{ccc} 
   S & \ell & K\\ 
   M_S & m_\ell & Q 
\end{array}\right) 
\left(\begin{array}{ccc} 
   S' & \ell' & K'\\ 
   M'_S & m'_\ell & Q' 
\end{array}\right), 
\end{split}
\end{equation} 
 with
\begin{equation}\label{b-matrix} 
\begin{split}
 b^{\ell K}_{\ell' K'}(K'') =& {2\pi\over k} \sum_{J_tjj'} 
 T^{J_t}_{n,n'}(E) 
(-1)^{J_t+J +S}\sqrt{(2j+1)(2j'+1)} 
\\ 
&\times (2J_t+1) 
\left\lbrace\begin{array}{ccc} 
   \ell & J_t & j\\ 
   J & S & K 
\end{array}\right\rbrace 
\left\lbrace\begin{array}{ccc} 
   \ell' & J_t & j'\\ 
   J' & S' & K' 
\end{array}\right\rbrace 
\left\lbrace\begin{array}{ccc} 
   J' & K' & J_t\\ 
   K  & J  & K'' 
\end{array}\right\rbrace 
\end{split}
\end{equation}

Introducing these last expressions in Eq.~(\ref{isotropic-cross-section}), and considering
no coherent terms, $i.e.$ $M_1=M_2\equiv M $ and $M'_1=M'_2\equiv M'$, the inelastic
cross section becomes
\begin{eqnarray}\label{total-cross-section} 
\sigma^0(\alpha J M; \alpha' J' M') &=& 
 \sum_K (-1)^{J+J' -M -M'} (2K+1) 
\left(\begin{array}{ccc} 
   J & J & K\\ 
   M & -M & 0 
\end{array}\right) 
\left(\begin{array}{ccc} 
   J' & J' & K\\ 
   M' & -M' & 0 
\end{array}\right) 
\quad \sigma^{K}_{\alpha J\rightarrow \alpha' J'}(E),  
\end{eqnarray} 
where the state multipoles are given by \citep{Kerkeni+00,Kerkeni02} 
\begin{eqnarray}\label{cross-section-state-multipoles} 
 \sigma^{K}_{\alpha J\rightarrow \alpha' J'}(E) = \sum_{K'} (-1)^{J+J'+K+K'} (2K'+1) 
\left\lbrace\begin{array}{ccc} 
   J' & J & K'\\ 
   J & J' & K 
\end{array}\right\rbrace 
\quad B(J,J'; K').
\end{eqnarray}
The rate constant state multipoles of Eq.~(\ref{rate-multipoles-inverse})
are directly obtained as
\begin{eqnarray}\label{rates-K}
C^{(K)}( \alpha' J'\leftarrow \alpha J )= \sqrt{\frac{2J'+1}{2J+1}}
N_{H^\circ}   \int v^2 dv f(T,v)\,
 \sigma^{K}_{\alpha J\rightarrow \alpha' J'}(E).
\end{eqnarray} 

The $B(J,J'; K')$ quantities are a generalization of the Grawert factors defined  
as \citep{Kerkeni+00,Kerkeni02}  
\begin{eqnarray}\label{Grawert-factors} 
 B(J,J'; K') = {1\over 4\pi (2S+1) } \sum_{\ell \ell'}\sum_{NN'} 
(2N+1) (2N'+1)\quad \left\vert b^{\ell N}_{\ell'N'}(K')\right\vert^2. 
\end{eqnarray} 
Note that the multipole expansion of the cross section,
 Eq.~(\ref{total-cross-section}),
presents the same symmetry properties of that of the rates of 
Eq.~(\ref{rate-multipoles}),
associated to the spherical symmetry introduced by the integration
over an isotropic distribution of ${\hat k}$ in Eq.~(\ref{isotropic-cross-section}).

\subsection{Calculation of S-matrix} 

Until here, the  treatment is general to any atom A, and the full
problem has been formalized in terms of the S-matrix elements. In order to evaluate
it, we should then particularize to the problem under study, which considers
atoms A with no nuclear spin colliding with an Hydrogen atom.
 The total Hamiltonian of the system
is given by

\begin{eqnarray}\label{total-Hamiltonian} 
H = -{\hbar^2\over 2\mu }\left({2\over R} {\partial \over \partial R}+ {\partial ^2\over \partial R^2}\right) 
+ {\bmath{\ell}  ^2\over 2\mu R^2} + H_{el} + H_{SO}^A, 
\end{eqnarray} 
where $\boldsymbol{\ell}$ is 
the end-over-end orbital angular momentum associated to the 
internuclear distance ${\bf R}$. 

The total wave
function is expanded as
\begin{eqnarray}\label{total-wvf} 
\left\vert \Psi^{J_t M_t n}_E\right\rangle 
=\sum_{n'}{ \Phi^{J_t M_t n}_{n'}(R; E)\over R} 
  \,\vert {\cal Y}^{J_t M_t}_{n'}\rangle. 
\end{eqnarray}
where ${\cal Y}^{J_t M_t}_n$ are eigenfunctions of the total angular
momentum,  with eigenvalue $J_t$, defined in a space-fixed frame
with the z-axis parallel to the observation direction as
\begin{eqnarray}\label{total-sf-functions} 
\vert {\cal Y}^{J_tM_t}_{n}\rangle 
= \sum_{m_\ell m} (-1)^{j-\ell + M_t} \sqrt{2J_t+1} 
      \left(\begin{array}{ccc} 
             j & \ell & J_t\\ 
             m & m_\ell & -M_t 
      \end{array}\right) 
Y_{\ell m_\ell}(\theta,\phi)  
 \vert j m ; \alpha J S\rangle, 
\end{eqnarray} 
where $m_\ell$, $m$ and $M_t$ are the projections of 
$\boldsymbol{\ell} $, ${\boldsymbol j}$ and ${\boldsymbol J}_t$ 
on the space-fixed $z$-axis, respectively. 
In this expression, the wavefunctions of the A($^{2S_A+1}L_{J}$)+H($^{2S+1}S$) fragments 
are expressed as 
\begin{eqnarray}\label{fragments-functions} 
  \vert j m ; \alpha J,  S \rangle = 
\sum_{M, M_S} (-1)^{J-S+m} \sqrt{2j+1} 
      \left(\begin{array}{ccc} 
             J & S & j\\ 
             M & M_S & -m 
      \end{array}\right) 
    \vert J M ; L, S_A \rangle  
\vert S M_S\rangle. 
\end{eqnarray} 
where $\vert S M_S\rangle$ are the eigenfunctions of   H,
and the A($^{2S_A+1}L_J$)  atoms are described by the states
\begin{eqnarray}\label{A-functions} 
  \vert J M ; \alpha\equiv L, S_A \rangle = 
\sum_{M_L, M_A} (-1)^{L-S_A+M} \sqrt{2J+1} 
      \left(\begin{array}{ccc} 
             L & S_A & J\\ 
             M_L & M_A & -M
      \end{array}\right) 
    \vert\varphi_{LM_L}\rangle \vert S_A M_A\rangle 
\end{eqnarray}
where  ${\boldsymbol J}= {\boldsymbol L} 
+ {\boldsymbol S}_A$ is the total angular momentum of A ($ {\boldsymbol L} $ 
is the orbital electronic part, described by the functions 
$\vert\varphi_{LM_L}\rangle$; $ {\boldsymbol S}_A$ the spin of A, 
described by the $\vert S_A M_A\rangle$ functions).

Introducing Equation~(\ref{total-wvf}) into the Schr\"odinger equation, 
multiplying by $\langle{\cal Y}^{J_tM_t}_{j\ell; \alpha J}|$, and integrating 
over electronic and angular variables, the following system of coupled 
differential equations result:
\begin{eqnarray}\label{close-coupling} 
\left\lbrace -{\hbar^2\over 2\mu}{\partial^2\over \partial R^2} 
            +{\hbar^2\ell(\ell+1)\over 2\mu R^2} + E^A_{LS, J} - E 
\right\rbrace\Phi^{J_tM_t n }_{n}(R;E) 
= -\sum_{n'} 
\left\langle {\cal Y}^{J_tM_t}_{n} \left\vert {\cal H}_{el} 
   \right\vert {\cal Y}^{J_tM_t}_{n'}\right\rangle 
\Phi^{JM\ell n}_{n'}(R;E).\nonumber\\ 
\end{eqnarray} 
These close-coupling equations are solved numerically using a Fox-Goodwin-Numerov 
method \citep{Gadea-etal:97}, subject to the usual boundary conditions:
\begin{eqnarray}\label{asymptotic-conditions} 
\Phi^{J_tM_t n}_{n'}(R\rightarrow\infty; E) 
\propto \sqrt{ {\mu\over 2  \pi\hbar^2}}\left\lbrace 
  \delta_{nn'} 
  {e^{-i( k R-\ell\pi/2)}\over \sqrt{k}}  
- S^{J_{t}}_{nn'}(E)  {e^{i( k' R-\ell'\pi/2)}\over \sqrt{k'}}  
\right\rbrace, 
\end{eqnarray} 
from where the scattering matrix $ S^{J_t}_{nn'}(E)$ is obtained.

The resolution of the close-coupling equations,
Eq.~(\ref{close-coupling}),
requires the evaluation  of the electronic Hamiltonian matrix elements
described below.

\subsection{Electronic matrix elements and approximations} 

The space-fixed functions of Equation~(\ref{total-sf-functions}) 
are eigenfunctions of $\boldsymbol{\ell}^2$ with eigenvalues 
$\hbar^2 \ell(\ell+1)$. 
However, the matrix elements of the electronic terms of the Hamiltonian, $H_{el}$ and $H^A_{SO}$,
deserve some comments. For treating these two terms we are considering three
approximations, which are the commonly used in the previous quantum treatments of collisional
depolarization of atoms \citep{Kerkeni+00,Kerkeni02}:

\begin{enumerate}

\item First, the spin-orbit is considered as constant  as a function of the internuclear distance $R$
and is entirely due to A atom. It is therefore 
convenient to consider  the isolated Hamiltonian of the A atom, 
\begin{eqnarray}\label{Hamiltonian-A} 
H^A = H_{el}^A + H_{SO}^A, 
\end{eqnarray} 
whose eigenfunctions are those of Equation~(\ref{A-functions}), with  
eigenvalues  $E^A_{LS_A, J}$. Here, we shall use the experimental values  
obtained from NIST. For this reason, 
 the non-relativistic electronic Hamiltonian is partitioned as 
\begin{eqnarray}\label{electronic-Hamiltonian} 
H_{el} = H_{el}^A + {\cal H}_{el}.
\end{eqnarray}

\item Second, we limit the states of A to a particular $\alpha\equiv L S_A$  subspace, 
thus not allowing transitions between different $L$ values. Thus, $H_{el}^A $
is equal in each $\alpha$ subspace. We denote the eigenvalues of 
$H_{el}$ and ${\cal H}_{el}$ as
$V^{S_t}_{L\Lambda}(R)$
and ${\cal V}^{S_t}_{L\Lambda}(R)$, respectively, 
with
 ${\cal V}_{L\Lambda}^{S_t}(R) = V_{L\Lambda}^{S_t}(R) -E_{L\Lambda}^{S_t}$, obtained 
for each internuclear distance $R$ 
and with $E_{L\Lambda}^{S_t}= V_{L\Lambda}^{S_t}(R\rightarrow\infty) $. 

\item Finally, the third is the usual Born-Oppenheimer approximation solving the non-relativistic 
 electronic equation 
\begin{eqnarray}\label{electronic-equation} 
H_{el} \vert\varphi_{L\Lambda}^{S_t}\rangle \vert S_t \Sigma; S_A S\rangle = 
 V_{L\Lambda}^{S_t}(R) \vert\varphi_{L\Lambda}^{S_t}\rangle \vert S_t \Sigma; S_A S\rangle. 
\end{eqnarray} 
These calculations are performed in a body-fixed frame, with the z-axis along 
the internuclear vector ${\bf R}$, related by a rotation  to the space-fixed frame used in the
whole treatment described above. In this new frame, the projections are denoted by greek 
letters, $\Lambda$ being the projection of ${\bf L}$ on the body-fixed z-axis, while
$M_L$ is its projection of the space-fixed z-axis.  In addition, 
in this approximation  $L$  is not a good quantum number  due to the cylindrical symmetry
of the problem, but it will be considered as constant since the orbital angular momentum
of hydrogen is zero.  Finally, the calculations are done for a total spin  ${\boldsymbol S}_t= 
{\boldsymbol S}_A+{\boldsymbol S}$. These $ V_{L\Lambda}^{S_t}(R) $ eigenvalues are described 
to the $\left\langle {\cal Y}^{J_t M_t}_n \vert {\cal H}_{el} \vert {\cal Y}^{J_t M_t}_{n'}\right\rangle$ matrix elements
of the present treatment as described below.

\end{enumerate}

For doing this transformation we shall define
 the functions for the total spin in the body-fixed frame as 
\begin{eqnarray}\label{total-spin-functions} 
  \vert S_t \Sigma_t ; S_A, S \rangle = 
\sum_{\Sigma_A, \sigma} (-1)^{S_A-S+\Sigma_t} \sqrt{2S_t+1} 
      \left(\begin{array}{ccc} 
             S_A & S & S_t\\ 
             \Sigma_A & \sigma & -\Sigma_t 
      \end{array}\right) 
    \vert S_A \Sigma_A \rangle \vert S \sigma\rangle, 
\end{eqnarray} 
and the functions of the fragments, Eq.~(\ref{fragments-functions}), in the body-fixed frame
  as
\begin{eqnarray}\label{fragments-functions-total-spin} 
  \vert j \Omega ; J, L, S_A, S \rangle &=& 
\sum_{S_t \Sigma_t \Lambda} (-1)^{-S_t-S_A-S-j+\Omega} \sqrt{(2j+1)(2J+1)(2S_t+1)} \\ 
& &\times      \left\lbrace\begin{array}{ccc} 
             S_t & S_A & S\\ 
             J & j & L 
      \end{array}\right\rbrace 
      \left(\begin{array}{ccc} 
             j & S_t & L\\ 
             \Omega & -\Sigma_t & -\Lambda 
      \end{array}\right) 
     \vert\varphi_{L\Lambda}^{S_t}\rangle  \vert S_t \Sigma_t; S_A S\rangle, 
\nonumber 
\end{eqnarray} 
where the functions $\vert\varphi_{L\Lambda}^{S_t}\rangle$ are the functions
obtained in Eq.~(\ref{electronic-equation}).
Finally, the eigenfunctions of well defined total angular momentum
in the body-fixed frame are expressed as
\begin{eqnarray}\label{total-bf-functions} 
\vert W^{J_tM_t}_{j\Omega; \alpha}\rangle 
= \sqrt{2J+1\over 4\pi} D^{J_t*}_{M_t\Omega}(\phi,\theta,0)  
 \vert j \Omega ; \alpha J S \rangle, 
\end{eqnarray} 
where $\theta,\phi$ are the polar angles of ${\boldsymbol R}$; 
$D^{J*}_{M\Omega}$ are rotation Wigner functions \citep{Zare-book} 
corresponding to the $M$ and $\Omega$ projections on the z-axis of the 
space-fixed and body-fixed frames, respectively.

Using Equations~(\ref{fragments-functions-total-spin})-(\ref{total-bf-functions}),
the ${\cal H}_{el}$ matrix elements in the body-fixed representation 
take the form 
\begin{equation}\label{bf-Hel-elements} 
\begin{split}
\left\langle W^{J_tM_t}_{j\Omega; \alpha} \left\vert {\cal H}_{el}\right\vert   W^{J_tM_t}_{j'\Omega'; \alpha'}\right\rangle 
=& \delta_{\Omega,\Omega'}  (-1)^{-S_A-S_A'-2S-j-j'+2\Omega}\sqrt{(2J+1)(2J'+1)(2j+1)(2j'+1)} 
\\ 
&\times  
\sum_{S_t} (-1)^{-2S_t} (2S_t+1)\delta_{S,S'} 
\left\lbrace\begin{array}{ccc} 
   S_t& S_A & S \\ 
   J & j & L 
\end{array}\right\rbrace 
\left\lbrace\begin{array}{ccc} 
   S_t& S'_A & S' \\ 
   J' & j' & L' 
\end{array}\right\rbrace 
 \\ 
&\times\sum_{\Lambda}\delta_{L,L'} {\cal V}_{L\Lambda}^{S_t}(R)\sum_{\Sigma_t} 
\left(\begin{array}{ccc} 
    j & S_t& L\\ 
   \Omega & -\Sigma_t & -\Lambda 
\end{array}\right) 
\left(\begin{array}{ccc} 
    j' & S_t& L\\ 
   \Omega' & -\Sigma_t & -\Lambda 
\end{array}\right). 
\end{split}
\end{equation} 
Using the transformations between
the space-fixed and body-fixed funtions,
\begin{subequations}\label{sf-bf-transformation} 
\begin{gather}
\vert {\cal Y}^{JM}_{j\ell; \alpha}\rangle 
 = \sum _\Omega \vert W^{JM}_{j\Omega; \alpha}\rangle  
(-1)^{j-\ell+\Omega}\sqrt{2\ell+1} 
  \left(\begin{array}{ccc} 
         j & \ell & J\\ 
   \Omega & 0 & -\Omega 
  \end{array}\right), 
\\ 
\vert W^{JM}_{j\Omega; \alpha}\rangle 
= \sum_\ell \vert {\cal Y}^{JM}_{j\ell; \alpha}\rangle  
(-1)^{\ell-j -\Omega}\sqrt{2\ell+1} 
  \left(\begin{array}{ccc} 
         j & \ell & J\\ 
   \Omega & 0 & -\Omega 
  \end{array}\right), 
\end{gather}
\end{subequations}
the electronic Hamiltonian in the space-fixed frame takes the form 
\begin{equation}\label{sf-Hel-elements} 
\begin{split}
\left\langle {\cal Y}^{JM}_{j\ell; \alpha} \left\vert {\cal H}_{el} 
   \right\vert {\cal Y}^{JM}_{j'\ell'; \alpha'}\right\rangle =& 
\sum_\Omega (-1)^{j+j'-\ell-\ell'+2\Omega}\sqrt{(2\ell+1)(2\ell'+1)}\\ 
&\times
\left(\begin{array}{ccc} 
       j &\ell & J\\ 
       \Omega & 0 & -\Omega 
\end{array}\right) 
\left(\begin{array}{ccc} 
       j' &\ell' & J\\ 
       \Omega & 0 & -\Omega 
\end{array}\right) 
\left\langle W^{JM}_{j\Omega; \alpha} \left\vert {\cal H}_{el}\right\vert   W^{JM}_{j'\Omega'; \alpha'}\right\rangle 
\nonumber, 
\end{split}
\end{equation} 
which is equivalent to those reported by \cite{Launay-Roueff:77b,Launay-Roueff:77a}
for some particular case.

\section{Quantum dynamical results}

An important feature of the atoms in the alkaline-earth 
series is the positionning of the two first excited states,
as discussed by \cite{Allouche-etal:92}. Whereas for Be~{\sc i} and Mg~{\sc i}, 
the lowest excited states are $^3P^\circ$ and $^1P^\circ$ corresponding to a $n s^1 n p^1 $ configuration, 
for Ca~{\sc i} and Sr~{\sc i} the low-lying $(n - 1)d$ orbital yields the lowest excited states 
$^3P^\circ$ and $^3D$,
corresponding to the $ns^1 np^1$ and $ns^1 (n-1)d^1$ configurations, respectively.
In Ba~{\sc i} the two first excited states, $^3D$ and $^1D$, correspond to the $6s^1 5d^1$ configuration.
The situation for cations, with a single valence electron, is analogous but simpler.
The importance of these arguments will be discussed below
and emphasizes the need
of a careful choice of the electronic basis to get accurate atomic energies for the excited states.

The electronic basis set used for magnesium and strontium are the all-electron, 
augmented correlation-consistent polarized core-valence 
 basis sets, aug-cc-pVTZ \citep{Woon-Dunning}. For calcium 
the  Def2-QZVP basis set was used, obtained from EMSL database \citep{Weigend-Ahlichs:05}.
For barium, 10 electron small-core
scalar relativistic effective core potentials, 
together with the corresponding valence basis sets \citep{Lim-etal:06}, ECP46MDF, were employed.
For hydrogen we used the augmented, correlation-consistent, polarized basis set, aug-cc-pVTZ,
of \cite{Dunning:89}.

\begin{figure}[t]
\includegraphics[scale=0.75]{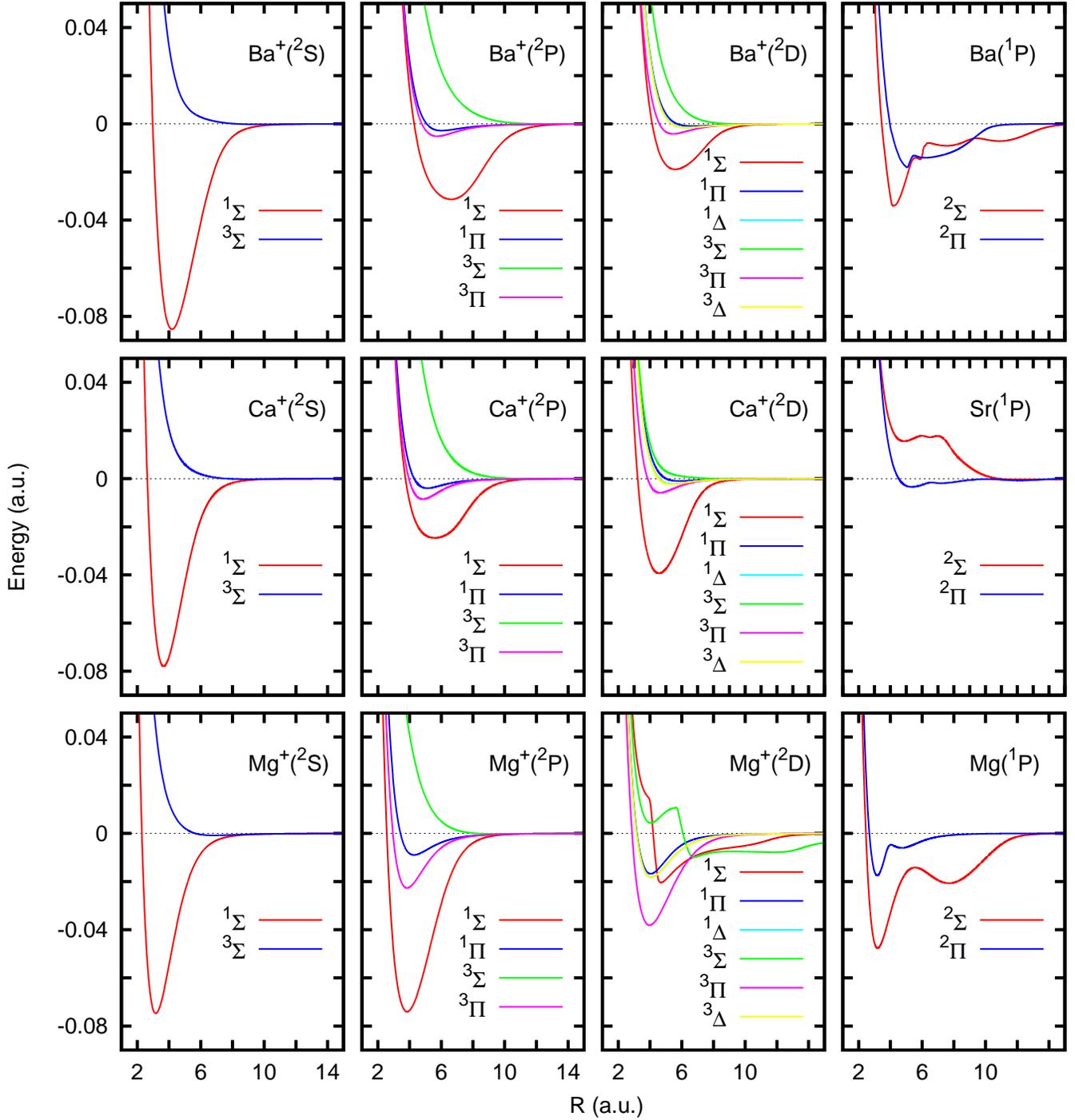}
\caption{\label{pes} Calculated ${\cal V}^{S_t}_{L\Lambda}(R)$ potential energy curves 
correlating to the atom A($^{2S_A+1}L$), as indicated in each panel. 
In all cases the zero of energy is at the non relativistic energy
of atoms at infinite distance. 
}
\end{figure}

The electronic adiabatic potentials, ${\cal V}^S_{L\Lambda}(R)$, were
calculated in two steps. First, the molecular orbitals were obtained
using a complete active space self consistent field (CASSCF) method,
with an active space including $ns, np, (n-1)s, (n-1)p, (n-1)d$ orbitals
of the alkaline-earth atom, and the $1s$ of hydrogen.   
In some cases, additional $s,p,d$ orbitals were introduced 
to properly account for $A^+-H^-$ ionic states and get the desired degeneracy
in the different atomic asymptotes. 
Second, with these reference states, a
multi-reference configuration interaction (MRCI) method \citep{Werner-Knowles:88}, was used to 
include the electronic correlation. All these calculations were performed
with the MOLPRO package \citep{MOLPRO-package}. Some potential curves are shown in Fig.~\ref{pes}.

The close coupled (CC) equations~(\ref{close-coupling}) were solved for each $LS_A$ term,
including all possible $J$-levels, $\vert L-S_A\vert \leq J \leq L+ S_A$, for all
the systems considered in this work. 
The differential equations were integrated
using a spatial grid of 20000 points in the interval $1\leq R \leq 50$ a.u.
The ${\cal  V}^S_{L\Lambda}(R)$ {\it ab initio} points, calculated 
 in a considerably coarser grid, were interpolated using cubic splines.
The CC equations were integrated for 1000 energies in the interval $E$=10 to 50000 cm$^{-1}$, 
and for each total angular momentum $J^t$, $0\leq J^t\leq J_{\rm max}$, where
$J_{max}$ was determined as that for which the S-matrix becomes diagonal at $E$=50000 cm$^{-1}$.
Depending on the system and state considered, $J_{\rm max}$ was between 500 and 1500.
The S-matrices were then used to calculate $\sigma^{K}_{\alpha J\rightarrow \alpha' J'}(E)$ 
according to equation~(\ref{cross-section-state-multipoles}). 
Finally, the state-multipoles of the rates, given in Eq.~(\ref{rate-multipoles}),
were calculated by integrating numerically over a Maxwell-Boltzman distribution,
 Eq.~(\ref{rates}),
using the energy grid described above.
From them the total inelastic rates, Eq.~(\ref{inelastic-rate-moments}),
 and the depolarization rates, Eq.~(\ref{depolarizing-rate}), were obtained.
 
{\bf The rates $\bar{C}(\alpha J)$ (see Eq. \ref{inelastic-rate-moments})
and $g^K(\alpha J)$ (see Eq. \ref{depolarizing-rate})
were fitted using the following simple analytical functions ($T$ being the temperature and $N_H$ the neutral hydrogen number density): }

\begin{equation*}
\bar{C}(\alpha J)=a_0 \times 10^{-9}\left(\frac{T}{5000}\right)^{b_0} N_{H}, \quad 
g^{K}=a_K \times 10^{-9}\left(\frac{T}{5000}\right)^{b_K} N_{H},
\end{equation*}
where $N_H$ is the neutral hydrogen number density (in cm$^{-3}$), and 
the $a_K$ (s$^{-1}$cm$^3$) and $b_K$ coefficients are given in Table~\ref{depolarization-constants}.
All the $C^{(K)}( \alpha J\leftarrow \alpha' J')$ needed to solve Eq.(\ref{master-equation-densities-K}) are
listed in the appendix.
All the inelastic transitions correspond to the same $\alpha=L,S_A$ manifold, and therefore
do only exist when both $L$ and $S_A$ are different from zero and several $J$ states appear.
When there are only two $J$ values,
these rates fulfill the Einstein-Milne relation \citep{Innocenti:04}.


The cations present the simplest structure, with a closed shell core and a single active electron. 
The $a_1$ coefficient for the ground $^2S$ state increases with the size of the core, from Mg~{\sc ii}
to Ca~{\sc ii}, to Ba~{\sc ii}. 
This is because the larger the cation, the larger the number of partial waves required for convergence,
which yields larger cross sections, and inelastic and depolarizing rates. 
A similar trend is observed for $a_1$ in the excited $^2P^\circ$ states of the cations, even when the $^1\Sigma$
state correlating to Mg~{\sc ii} (3p, $^2P$) is about twice deeper than in Ca~{\sc ii} and Ba~{\sc ii} 
(Fig.~\ref{pes}).
The $a_0$ is significantly smaller for Ba~{\sc ii} ($^2P$) than for  Ca~{\sc ii} ($^2P$), while
the potential curves involved for these two systems are rather similar (Fig.~\ref{pes}).
The reason for this is the strong dependence ($\sim Z^4$) of the fine structure splitting with the  
atomic number $Z$ ---the energy separation  between the $J=1/2$ and $3/2$ levels is 
 91.6, 222.9, and 1690.8 cm$^{-1}$, for  Mg~{\sc ii},  Ca~{\sc ii} and Ba~{\sc ii}, respectively.

The behaviour of the $^2D$ states of the cations is strikingly different, 
$a_0$ and $a_1$ being $\sim$6 times larger for Mg~{\sc ii} than for Ca~{\sc ii} or Ba~{\sc ii}. 
The reason is that, while in Mg~{\sc ii} the $^2D$ states correspond to the fourth electronic term,
in Ca~{\sc ii} and Ba~{\sc ii} they belong to the second term. 
This makes the  $^1\Sigma$ and $^3\Sigma$ potential curves correlating 
to the Mg~{\sc ii} ($^2D$) cross with the ionic states of Mg~{\sc i} + H$^+$. 
The adiabatic potential energy curves used in the present Born-Oppenheimer quantum {\it ab initio} 
approximation are more involved as a consequence of the avoided 
crossings \citep{Khemiri-etal:13}. 
This behaviour is the responsible of the anomaly of the  Mg~{\sc ii} ($^2D$) levels.

The neutral alkaline-earth atoms are more complex, since they have a closed shell core surrounded
by two electrons, and there are many other ionic states correlating to the different electronic terms
of the cations which may cross with the excited electronic states. 
The ground state is always an isotropic $^1S$ state,
and the main transition is towards the $^1P^\circ$ state. 
For heavier atoms, the $^1P^\circ$ state
produces a progressively more complex structure, due to avoided crossings with ionic and low-lying covalent states. 
For Mg~{\sc i} ($^1P^\circ$) this crossing produce a double well in the $^2\Sigma$ and $^2\Pi$ states with a 
rather long interaction length (Fig.~\ref{pes}).
The $a_i$ values obtained in this work are very close to those reported by \cite{Kerkeni02}, within 1-2 $\%$,
showing that the potential curves are rather similar.

The  potential energy curves correlating to Sr~{\sc i} ($^1P^\circ$) are very close to those correlating to
$^1D$ state. 
A precise description of this situation is difficult 
and the results of the calculations become rather sensitive to the electronic basis
used and the method applied. 
This may be the reason why the present results differ from those of \cite{Kerkeni+03} in about 10-15$\%$.
For Ba~{\sc i} the situation is even worse because its adiabatic curves are the result of many avoided
crossings between different covalent and ionic states \citep{Allouche-etal:92}, and
the $^2\Sigma$ and $^2\Pi$ potential energy curves
present a very complex structure (Fig.~\ref{pes}). 
Due to the long range of the interaction, the number of partial waves required to get convergence 
increases noticeably, producing a significant increase of the inelastic and depolarizing rates, 
which are far larger than in the other cases studied here.

In the quantum {\it ab initio} approach used here and most
previous studies \citep{Kerkeni02,Kerkeni+03}, an adiabatic
description of the electronic states is used. In this description, 
the states correlating to given states of the $A(\alpha, J)+ H$
fragments change adiabatically as they cross with others, essentially of
ionic character. This change is consistent with the commonly used
Born-Oppenheimer approximation, in which it is assumed that the electrons move much faster than the nuclei,
which allows an instantaneous transition from a covalent  $A(\alpha, J)+ H$ state
to  ionic  $A^+(\alpha', J')+ H^-$ or $A^-(\alpha'', J'')+H^+$ states, 
at the precise distance where their energies coincide.

The opposite (diabatic) situation is the one
in which this covalent/ionic transitions are neglected, considering
that the $A(\alpha, J)+ H$ character is preserved for all the
internuclear distances. Such description is the one used in the 
semiclassical approach of 
\cite{Brueckner:71,Anstee-Omara:95,Derouich+03}.
This different treatment of the interaction potential 
may explain the dissagrement found for the depolarization
rates obtained for Sr~{\sc I} 460.7~nm line
 between the adiabatic quantum {\it ab initio} method 
in Table~\ref{depolarization-constants}
and the diabatic semiclassical method of \cite{Faurobert+95}. For example,
our results for Mg~{\sc ii}($^2P$) are very close to the
ones by \cite{Kerkeni02} (also adiabatic), but the semiclassical 
results of \cite{Derouich+03} are larger by about a $10\%$.

The adiabatic description is better suited for low collision energies
while the diabatic one is better at high energies.
Actually, a combined description is needed, in which transitions between covalent 
and ionic states are allowed by including the non-adiabatic couplings,
thus allowing transitions among different $L,S_A$ manifolds. 
In order to accomplish this, it is convenient to change from the adiabatic
representation, where the couplings show very
sharp variations at the crossings, to a diabatic representation that includes the
 couplings among the different states.
Such diabatization procedure has been already done for some of the states
of MgH \citep{Belayev-etal:12} and CaH$^+$ \citep{Habli-etal:11}. 
Some work in this direction is now being done in order to assess the necessity of a more exact 
treatment for the collisional depolarization of atoms in excited electronic
states including inelastic transitions between different $L, S_A$ manifolds.

\begin{table}[t]
\caption{\label{depolarization-constants}
$\bar{C}(\alpha J)=a_0 10^{-9} (\frac{T}{5000})^{b_0}N_H$
and 
$g^{K}=a_K 10^{-9} (\frac{T}{5000})^{b_K}N_H$ with $K>0$.$^e$}
\begin{center}
\begin{tabular}{ccccccccccccccc}
\hline\hline
term  & $J$& {E(cm$^{-1}$)$^a$} & {$a_0$}  &  $b_0$  &  $a_1$  & $b_1$  & $a_2$  & $b_2$  & $a_3$  & $b_3$  & $a_4$  & $b_4$ & $a_5$  & $b_5$ 
\\
\hline
\multicolumn{15}{c}{Mg II(${}^2S$, ${}^2P$,${}^2D$)+H} \\
\hline
$3s{,}^2S$ & 1/2 &     0.0  & \multicolumn{1}{c}{--} & -- & 2.9704 & 0.36 & & & & & & & & \\
$3p{,}^2P^\circ$ & 1/2 & 35669.3  & 4.1553 & 0.391 & 5.8448 & 0.364 & & & & & & & \\
          & 3/2 & 35760.9  & 2.1505 & 0.354 & 5.2791 & 0.344 & 6.1487 & 0.360 & 5.8910 & 0.366 & & & & \\
$4s{,}^2S$ & 1/2 & 69804.9  & \multicolumn{1}{c}{--} & -- & 7.2124 & 0.34 & & & & & & & & \\
$3d{,}^2D$ & 5/2 & 71490.2  & 13.4215 & 0.345 & 26.1186 & 0.349 & 31.7442 & 0.3523 & 33.6754 & 0.359 & 35.9476 & 0.360  &  33.8079 & 0.350 \\
          & 3/2 & 71491.1  & 20.1389 & 0.345 & 30.3626 & 0.354 & 34.7140 & 0.360 & 33.9266 & 0.358 & & &  & \\
\hline
\multicolumn{15}{c}{Ca II(${}^2S$, ${}^2P$,${}^2D$)+H} \\
\hline
$4s{,}^2S$ & 1/2 &     0.0  & \multicolumn{1}{c}{--}  & -- & 3.6272     & 0.40      & & & & & &  & & \\
          &     &          & \multicolumn{1}{c}{--}  & -- & {\em 3.2935}$^b$ & {\em 0.451}$^b$ & & & & & &  & & \\
$3d{,}^2D$ & 3/2 &  13650.2 & 2.3294 & 0.333 & 3.9194 & 0.314 & 4.3988 & 0.310 & 4.1986 & 0.3124 & &  & & \\
          &     &          &        &       & {\em 3.3700}$^b$ & {\em 0.381}$^b$ & {\em 3.8120}$^b$ & {\em 0.376}$^b$ & {\em 3.6154}$^b$ & {\em 0.380}$^b$ & & & & \\
          & 5/2 &  13710.9 & 1.5892 & 0.307 & 3.4631 & 0.317 & 4.0100 & 0.299 & 4.4144 & 0.3015 & 4.5694 & 0.307& 4.1522 & 0.314 \\
          &     &          &        &       & {\em 2.9078}$^b$ & {\em 0.385}$^b$ & {\em 3.2387}$^b$ & {\em 0.34}$^b$1 & {\em 3.6143}$^b$ & {\em 0.359}$^b$ & {\em 3.7318}$^b$ & {\em 0.353}$^b$ & {\em 3.5506}$^b$ & {\em 0.387}$^b$ \\
$4p{,}^2P^\circ$ & 1/2 &  25191.5 & 4.3445 & 0.5344 & 6.8611 & 0.401 & & & & & &  & & \\
          &     &          &        &        & {\em 6.2873}$^b$ & {\em 0.476}$^b$ & & & & & &  & & \\
          & 3/2 &  25414.4 & 2.3594 & 0.450 & 6.5318 & 0.350 & 7.6949 & 0.378 & 7.2009 & 0.395 & &  & & \\
          &     &          &        &       & {\em 6.0059}$^b$ & {\em 0.406}$^b$ & {\em 7.0291}$^b$ & {\em 0.428}$^b$ & {\em 6.6772}$^b$ & {\em 0.418}$^b$ & &  & & \\
\hline
\multicolumn{15}{c}{Ba II(${}^2S$, ${}^2P$,${}^2D$)+H $^d$} \\
\hline
$6s{,}^2S$ & 1/2 &      0.0 & \multicolumn{1}{c}{--} & -- & 4.5783 & 0.39 & & & & & &  & & \\
$5d{,}^2D$ & 3/2 &   4873.9 & 1.3053  & 0.941 & 4.3374 & 0.376 & 5.1832 & 0.360 & 5.3721 & 0.3376 & &  & & \\
          & 5/2 &   5674.8 & 1.1497  & 0.6943 & 4.3606 & 0.343 & 4.9580 & 0.340 & 5.6322 & 0.336 & 5.9298 & 0.323  & 5.1728 & 0.345 \\
$6p{,}^2P^\circ$ & 1/2 &  20261.6 & 0.5658  & 1.264 & 9.6737 & 0.386 & & & & & &  & & \\
          & 3/2 &  21952.4 & 0.5153  & 0.708 & 8.4584  & 0.379 & 10.2608 & 0.413 & 9.6008 & 0.399& &  & & \\
\hline
\multicolumn{15}{c}{Mg I(${}^1P$, ${}^3S$, ${}^3P$)+H} \\
\hline
$3s3p{,}^3P^\circ$ & 0  & 21850.4 & 6.4795 & 0.413 &    &        &  &  & & & &  & & \\
            & 1  & 21870.5 & 4.8804 & 0.411 & 6.7143 & 0.411 & 7.0615 & 0.408 & & & &  & & \\
            & 2  & 21911.2 & 3.2480 & 0.402 & 4.9124 & 0.395 & 6.7226 & 0.401 & 7.1756 & 0.409 & 7.2994 & 0.413  & & \\
$3s3p{,}^1P^\circ$ & 1  & 35051.3 & \multicolumn{1}{c}{--} & -- & 13.696 & 0.415 & 11.944 & 0.437 & & & &  & & \\
            &    &         &    &    & {\em 13.836}$^c$ & {\em 0.329}$^c$ & {\em 10.79}$^c$ & {\em 0.367}$^c$ & & & &  & & \\
$3s4s{,}^3S$ & 1  & 41197.4 & \multicolumn{1}{c}{--} & -- & 6.1710 & 0.426 & 18.5106 & 0.426 & & & &  & & \\
\hline
\multicolumn{15}{c}{Sr I(${}^1P$)+H} \\
\hline
$5s5p{,}^1P^\circ$ & 1  &  21698.5     & \multicolumn{1}{c}{--}  & -- & 10.5432 & 0.30 & 8.42791 & 0.38 & & & &  & & \\
            &    &              &     &    & {\em 8.9875}$^c$ & {\em 0.376}$^c$ & {\em 8.0015}$^c$ & {\em 0.386}$^c$ & & & &  & & \\
\hline
\multicolumn{15}{c}{Ba I(${}^1P$,${}^1D$, ${}^3P$, ${}^3D$)+H  $^d$}   \\
\hline
$6s5d{,}^3D$ & 1  &   9034.0     & 5.6561 & 0.5814 & 8.6534 & 0.3938 & 10.6750 & 0.3577 & & & & & &  \\
            & 2  &   9215.5     & 5.6386  & 0.5683 & 8.7799  & 0.4296 & 9.1290 & 0.4222  &9.6180 &0.4161 & 10.4261 &0.4008  & & \\
            & 3  &   9596.5     &3.4702  &0.5797 & 5.3034 & 0.4543 &6.6440  & 0.4378 &7.6926 &0.4198 &8.7713 &0.4045  &9.8750 & 0.3655\\
$6s5d{,}^1D$ & 2  &  11395.4     & -- & -- &  9.5977 & 0.4558  &  10.0612 & 0.4483  & 11.2425 &0.4346 &  9.9601 & 0.4486 & & \\
$6s6p{,}^3P^\circ$ & 0  &  12266.0     &  0.2595 & 0.8466 &  &  &  &  & & & & & & \\
            & 1  &  12636.6     & 3.4062 & 0.8594 & 9.6736 & 0.3252 &12.1638  & 0.3022  & & & &  & & \\
            & 2  &  13514.7     & 1.3572 & 0.9596 & 7.6312 & 0.3774  & 11.4646  & 0.3842 & 12.5007 &0.3975 &13.3332 & 0.4126 & & \\
$6s6p{,}^1P^\circ$ & 1  &  18060.3     & \multicolumn{1}{c}{--} & -- & 18.5210  & 0.4240  & 16.3480  & 0.424  & & & & &  \\
\hline\hline
\end{tabular}
\end{center}
$^a$ From the NIST Atomic Spectra Database {\tt http://physics.nist.gov/asd} \citep{Kramida+13} 

$^b$ From~\cite{Kerkeni+03}

$^c$ From~\cite{Kerkeni02}

$^d$ For Ba the isotopes 136 and 138 with zero nuclear spin have been considered, and they
give undistinguisable results.

$^e$ The $a_K$ coefficients are given in s$^{-1}$cm$^3$
\end{table}


\section{Depolarizing collisions in the solar atmosphere}

Now, we study the effect of the calculated depolarizing elastic collisions 
on the formation of resonance polarization patterns in the solar atmosphere.
In particular, we consider the resonant lines of Mg~{\sc i}, and Mg~{\sc ii}, 
Sr~{\sc i}, Ca~{\sc ii}, Ba~{\sc i}, and Ba~{\sc ii} listed in Table~\ref{atomic-lines}, 
and a semiempirical model of the quiet solar atmosphere 
such as the C model of Fontenla et al. \cite[][FAL-C]{FAL93}.

The generation of atomic polarization in atomic levels is detemined by the 
radiative rates involving those levels.
The radiation field in the solar atmosphere is quasi-thermal and relatively {\em weak},
the number of photons per mode being $\bar{n}\ll 1$.
For excited levels, radiative rates are therefore, dominated by spontaneous decay (no absorptions
to upper lying levels); in ground and metastable levels, the only possible
radiative rates are absorptions towards upper lying ones.
The mean-life time of an excited level $u$ is thus $\tau_{\rm life}=1/\sum_\ell A_{u\ell}$, 
where $A_{u\ell}$ is the Einstein coefficient for spontaneous emission 
in the $u\rightarrow \ell$ transition and the sum extends
over all $\ell$ levels radiatively connected to $u$; 
the mean-life time of the ground or metastable level $\ell$ is 
$\tau_{\rm life}=1/\sum_u B_{\ell u}J(\nu_{u\ell})$,
where $B_{\ell u}$ is the Einstein coefficient for absorption, 
$J=\int \phi_\nu d\nu\int\frac{d\Omega}{4\pi} I$
is the mean intensity (over the solid angle $\Omega$), integrated over the absorption profile $\phi_\nu$, 
and the sum extends over all the levels radiativelly connected to $\ell$.
Thus, consider the singly ionized akaline earths (see Figure~1). 
For the excited levels ${}^2P^\circ_{3/2}$,  
$\tau_{\rm life}^{-1}\approx A_k$ in Mg~{\sc ii}, 
$\tau_{\rm life}^{-1}\approx A_{K}+A_{849.8}+A_{854.2}$ in Ca~{\sc ii},
and $\tau_{\rm life}^{-1}\approx A_{455.4}+A_{585.3}+A_{614.1}$ in Ba~{\sc ii}. 
For the metastable levels, 
$\tau_{\rm life}^{-1}({}^2D_{3/2})=B_{866.2}J(866.2)+B_{849.8}J(849.8)$,
$\tau_{\rm life}^{-1}({}^2D_{5/2})=B_{854.2}J(854.2)$, in Ca~{\sc ii};
$\tau_{\rm life}^{-1}({}^2D_{3/2})=B_{649.6}J(649.6)+B_{585.3}J(585.3)$,
$\tau_{\rm life}^{-1}({}^2D_{5/2})=B_{614.1}J(614.1)$, in Ba~{\sc ii}.
Analogously, regarding the neutral alkaline earths (see Figure~1), 
for Mg~{\sc i}, $\tau_{\rm life}^{-1}({}^3S)=A_{b_1}+A_{b_2}+A_{b_3}$,
$\tau_{\rm life}^{-1}({}^3P^\circ_1)=B_{b_2}J(b_2)$, 
$\tau_{\rm life}^{-1}({}^3P^\circ_2)=B_{b_4}J(b_4)$,
and $\tau_{\rm life}^{-1}({}^1P^\circ)=A_{285.2}$; 
for Sr~{\sc i}, $\tau_{\rm life}^{-1}({}^1P^\circ)=A_{460.7}$;
for Ba~{\sc i}, $\tau_{\rm life}^{-1}({}^1P^\circ)=A_{460.7}$.

The coefficients $A_{u\ell}$ compiled from the NIST database are tabulated in Table \ref{atomic-lines},
from which the $B_{\ell u}=A_{u\ell}\frac{c^2}{2h\nu^3}g_u/g_\ell$ derive 
($c$ is the speed of light, $h$ the Planck constant, $\nu$ the frequency of the 
transition, and $g_i=2J_i+1$ the degeneracy of the level).
The mean intensity $J$ varies at each point in the atmosphere.
Deep in the atmosphere the radiation field is trapped and close to Planckian
$J\approx B_\nu$ ($B_\nu$ is the Planck function); in the upper layers of the atmosphere 
the radiation field may escape through the free boundary and $J$ 
strongly separates from $B_\nu$. 
The calculation of the actual values of $J$ requires computing the radiation field consistent
with the excitation state of the atoms in the atmosphere (non-LTE problem). 
The population of an atomic level $i$ may be expressed as 
\begin{equation}\label{population}
N_i={\cal N} \; 10^{A-12}\; \alpha \frac{g_i}{u(T)} \exp\left(-\frac{E_i}{k_{\rm B}T}\right) b_i,
\end{equation}
where ${\cal N}$ is the total number density of atoms, $A$ is the abundance of the element 
in the usual logarithmic scale in which $A_H=12$
 \citep[$A_{\rm Ca}=6.36$, $A_{\rm Mg}=7.58$, $A_{\rm Sr}=2.97$, $A_{\rm Ba}=2.13$;][]{Grevesse+96},
$\alpha$ is the fraction of the ionization state considered \citep[for simplicity, 
here we assumed it is given according to the Saha formula;][]{Mihalas78},
$E_i$ the excitation energy of the level, $k_{\rm B}$ the Boltzmann constant,
and $T$ the temperature, $g_i=2J_i+1$ the degeneracy of the level, 
$u(T)$ the partition function of the ion \citep[we used the tables of][]{Irwin81}, and 
$b_i$ the departure coefficient from a  purely LTE population.
The rigorous calculation of $J$ for the radiation field  and the $b_i$ 
for the atomic levels requires the solution of a set of non-linear, 
non-local, integro-differential equations ---the NLTE problem \citep[e.g., ][]{Mihalas78}.
We made the simple, rough estimate $b_i\approx 1$ for all the levels of interest,
from which the intensity may be obtained integrating the radiative transfer equation 
\begin{equation}\label{rte}
\frac{dI}{ds} = -\kappa I+\epsilon,
\end{equation}
where 
\begin{equation*}
\kappa=\frac{h\nu}{4\pi}B_{\ell u}N_\ell\phi_\nu, \quad {\rm and} \quad
\epsilon=\frac{h\nu}{4\pi}A_{u \ell}N_u\phi_\nu, 
\end{equation*}
are the absorption and emission coefficients, respectively, with
$N_\ell$ and $N_u$ being the population of the lower and upper level of the
transition and $\phi_\nu$ the Voigt absorption profile.
Considering the solar atmosphere as a plane-parallel atmosphere, 
the element of path along a ray is $ds=dz/\cos\theta$, with $z$ the height 
and $\theta$ the inclination angle of the ray with respect to the vertical.
Equation~(\ref{rte}) was numerically integrated using 
a short-characteristics scheme \citep{KunaszAuer88}; 
a Gaussian $N_\mu$-point quadrature was used for the angular integration 
and a trapezoidal $N_\nu$-point rule for the frequency integral 
(here, $N_\mu=21$ and $N_\nu=11$ were used).

\begin{figure}
\centering 
\includegraphics[scale=1,angle=0]{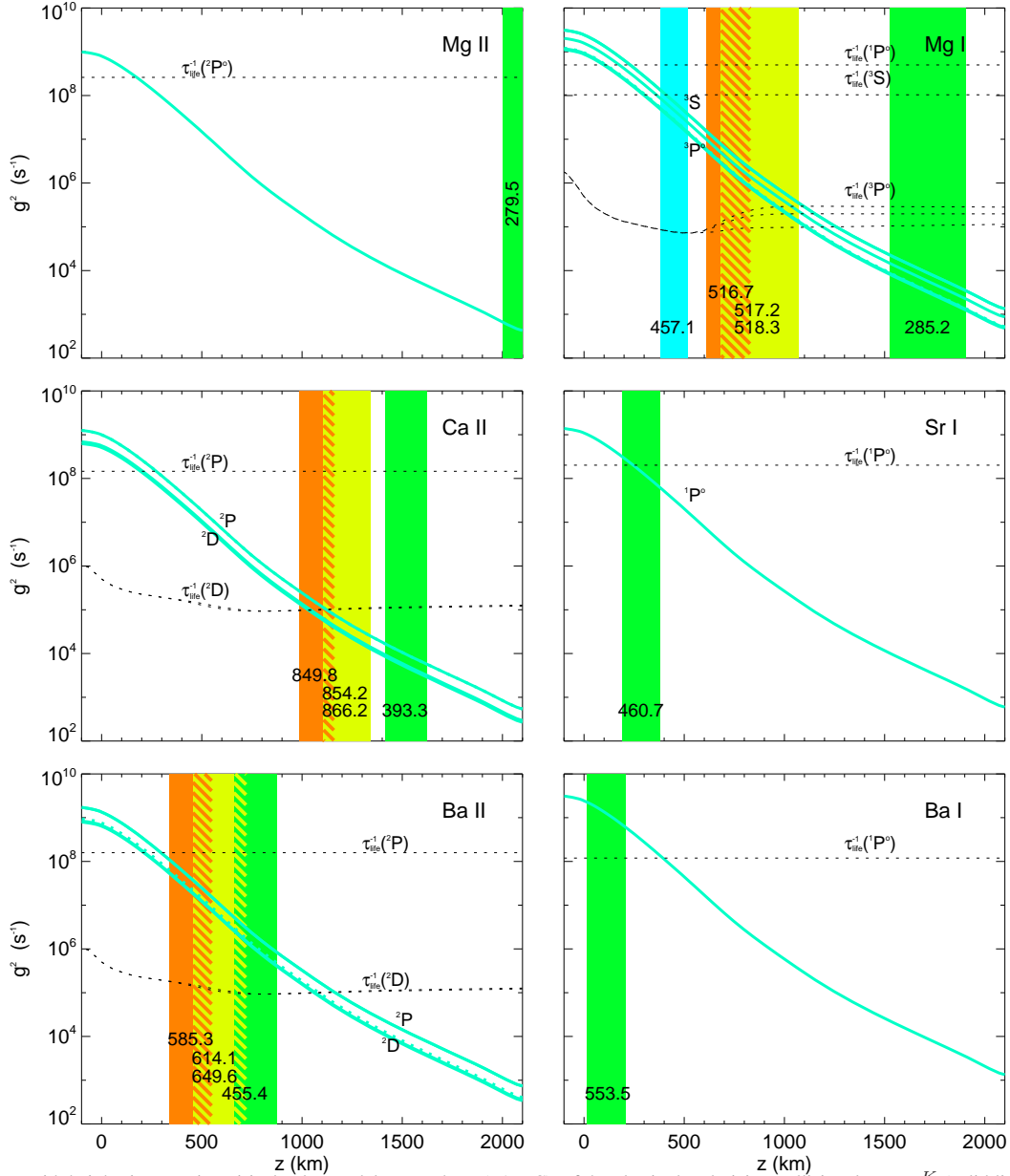} 
\caption{\label{sunny}
Variation with height, in a semiempirical solar model atmosphere 
(FAL-C), of the elastic depolarizing collisional rates $g^{K}$ (solid lines)
and radiative relaxation rates (dotted lines) in the lowest levels and
resonant transitions of Mg~{\sc ii}, Mg~{\sc i}, Ca~{\sc ii}, Sr~{\sc i},
Ba~{\sc ii}, and Ba~{\sc i}. 
The curves for $g^{2}$ and $g^{4}$ of the ${}^2D$ metastable levels of
Ca~{\sc ii} and Ba~{\sc ii}, and the ${}^3P$ levels of Mg~{\sc i}, are 
indistiguishable.
Colored bands show the estimated height of formation of the core 
of the resonant lines indicated (labels in nm), from disk center (left edge)
to $\mu=0.1$ (right edge).
}
\end{figure} 

\begin{table}
\caption{\label{atomic-lines} Atomic data for lines of astrophysical interest$^a$}
\begin{center}
\begin{tabular}{lccccc}
\hline \hline
$\lambda$ (nm) & lower term & $J_\ell$ & upper term & $J_u$ & $A_{u\ell}$ (s${}^{-1}$) \\
\hline
\multicolumn{6}{c}{Mg II} \\
\hline
279.553 ($k$) & $3s\; {}^2S$ & 1/2  & $3p\; {}^2P^\circ$ & 3/2 & $2.6\times 10^8$ \\
280.270 ($h$) &              & 1/2  &                  & 1/2 & $2.6\times 10^8$ \\
\hline
\multicolumn{6}{c}{Ca II} \\
\hline
393.366 ($K$) & $4s\; {}^2S$ & 1/2  & $4p\; {}^2P^\circ$ & 3/2 & $1.35\times 10^8$ \\
396.847 ($H$) &              & 1/2  &                  & 1/2 & $1.32\times 10^8$ \\
849.802 & $3d\; {}^2D$ & 3/2  & $4p\; {}^2P^\circ$ & 3/2 & $1.11\times 10^6$ \\
854.209 & & 5/2  & & 3/2 & $9.9\times 10^6$ \\
866.214 & & 3/2  & & 1/2 & $1.06\times 10^7$ \\
\hline
\multicolumn{6}{c}{Ba II} \\
\hline
455.403 & $6s\; {}^2S$ & 1/2  & $6p\; {}^2P^\circ$ & 3/2 & $1.11\times 10^8$ \\
493.407 &              & 1/2  &                  & 1/2 & $9.53\times 10^7$ \\
585.367 & $5d\; {}^2D$ & 3/2  & $6p\; {}^2P^\circ$ & 3/2 & $6\times 10^6$ \\
614.171 &              & 5/2  &                  & 3/2 & $4.12\times 10^7$ \\
649.690 &              & 3/2  &                  & 1/2 & $3.10\times 10^7$ \\
\hline
\multicolumn{6}{c}{Mg I} \\
\hline
285.212 & $3s^2\; {}^1S$ & 0  & $3s3p\; {}^1P^\circ$ & 1 & $5\times 10^8$ \\
457.109 & $3s^2\; {}^1S$ & 0  & $3s4s\; {}^3S$       & 1 & $2.54\times 10^2$ \\
516.732 ($b_1$)& $3s3p\; {}^3P^\circ$ & 0  & $3s4s\; {}^3S$ & 1 & $1.16\times 10^7$ \\
517.268 ($b_2$)&  & 1  &  & 1 & $3.46\times 10^7$ \\
518.360 ($b_4$)&  & 2  &  & 1 & $5.75\times 10^7$ \\
\hline
\multicolumn{6}{c}{Sr I} \\
\hline
460.733 & $5s^2\; {}^1S$ & 0  & $5s5p\; {}^1P^\circ$ & 1 & $2.01\times 10^8$ \\
\hline
\multicolumn{6}{c}{Ba I} \\
\hline
553.548 & $6s^2\; {}^1S$ & 0  & $6s6p\; {}^1P^\circ$ & 1 & $1.19\times 10^8$ \\
\hline
\hline
\end{tabular}
\end{center}
$^a${From the NIST Atomic Spectra Database {\tt http://physics.nist.gov/asd} (Kramida et al. 2013)}
\\
\end{table}

For resonance transitions at optical frequencies ($\nu$) 
and for the temperatures characteristic of the solar
atmosphere ($T\sim 6000$~K), 
$B_{\ell u}J\sim B_{\ell u} B_\nu(T)\sim A_{u\ell} \exp(-h\nu/K_{\rm B}T)$
($K_{\rm B}$ is the Boltzmann constant), i.e., $B_{\ell u}J\ll A_{u\ell}$ and hence,
the mean life time of ground and metastable levels ($\ell$) 
is a few orders of magnitude larger than the radiative life time
of excited levels $u$: $\tau_{\rm life}(\ell)\gg\tau_{\rm life}(u)$.

The variation with height of $g^{(K)}$ in the solar atmosphere 
is dominated by the exponential stratification of density, 
with just a minor correction from 
$T$ due to the weak dependence of the collisional rates ($b_K\sim 0.3$-0.4 according to  
Table~\ref{depolarization-constants}), 
falling six orders of magnitude from the bottom of the photosphere to 
the high chromosphere 2000~km above (see Figure~\ref{sunny}).

The role of depolarizing collisions on the formation of the scattering polarization 
patterns is determined by the relative importance of the radiative (polarizing)
and the collisional (depolarizing) rates at the height where the line {\em forms}.
We may estimate the {\em formation height}, $H$, of a spectral feature roughly as 
that for which the optical distance to the free surface (at $z_{\rm max}$)
is $\int_H^{z_{\rm max}} \kappa dz/\mu\approx 1$. 
Taking into account the definition of $\kappa$ (after Equation~(\ref{rte}))
and Equation~(\ref{population}):
\begin{equation}
\frac{c^2}{8\pi^{3/2}\nu^2} 10^{(A-12)}{\cal N}_0
\int_H^\infty 
\frac{A_{u\ell}}{\Delta\nu_D}
b_\ell
\frac{g_u \exp(-E_\ell/K_{\rm B}T)}{u(T)} \alpha \exp(-z/{\cal H}) \frac{dz}{\mu}\approx 1,
\end{equation}
where $\Delta\nu_D$ is the thermal Doppler width of the transition
and a strict exponential stratification ${\cal N}={\cal N}_0\exp(-z/{\cal H})$ 
(${\cal N}_0\approx 1.2\times 10^{17}$~cm$^{-3}$, ${\cal H}\approx 130$~km
for FAL-C), has been assumed for the total number density.
Figure~\ref{sunny} shows the regions where 
the core of several important alkaline-earth resonance lines form (vertical stripes) 
at different heliocentric distances, from disk center ($\cos\theta=1$; 
lowest part of the stripe), to $\cos\theta=0.1$ (upper part of the stripe)
characteristic of observations close to the solar limb.

The Mg~{\sc ii} $h$ and $k$-lines form very high in the chromosphere, just
short of the transition region. 
At those heights, due to the low density (and despite the temperature rise), 
the radiative rates of the upper level ${}^2P^\circ_{3/2}$ 
is several orders of magnitude larger than the collisional depolarizing rates.
The atomic polarization induced in that level will remain largely unaffected 
by collisions.

Calcium is 16 times less abundant than magnesium 
and the corresponding Ca~{\sc ii} $H$ and $K$-lines 
at 396 and 393~nm form $\sim$500~km lower in the chromosphere, 
but the radiative decay rate of the level ${}^2P^\circ_{3/2}$ 
is still orders of magnitude larger than $g^{K}$ and
its atomic polarization remains unaffected by depolarizing collisions.
The infrared triplet lines between the metastable ${}^2D$ levels
and the excited ${}^2P$ levels form lower in the chromosphere.
Interestingly, the $\tau_{\rm life}^{-1}({}^2D)$ is larger than 
$g^{K}$ in that region, which guarantees that the
atomic polarization generated in the metastable levels survives
to the depolarizing collisions.
Certainly this must be the case since it has long been shown 
that the origin of the observed scattering 
polarization pattern in this triplet \citep{Stenflo+00}, 
must be the differential absorption of light polarization components
(dichroism) due to the presence of a sizable amount of atomic polarization 
in the metastable levels \citep{MansoTrujillo03}.
Yet, $g^{K}$ is still of the order of the radiative rates, 
which means that the actual values of the observed polarization 
are modulated by the value of the collisions, and precise
values of $g^{K}$ are mandatory for the accurate diagnostic
of chromospheric magnetic fields via the Hanle effect in these
important lines.

The corresponding $D_{1, 2}$ lines of Ba~{\sc ii} (${}^2S-{}^2P$),
and the triplet (${}^2D-{}^2P$) form even deeper, in the region of the minimum of temperature 
(see Figure~\ref{sunny}).
The collisional rates are still unable to compete with the strong radiative 
rates involving the ${}^2P_{3/2}$ level, 
but they are high enough to completely depolarize the metastable
levels ${}^2D$.

{ Mg~{\sc i} is a minority species even in the region of the minimum of temperature
which makes it very sensitive to NLTE effects. In fact, our estimate for the 
ionization fraction $\alpha$ using the Saha-formula grossly overestimates the 
abundance of Mg~{\sc iii} in the upper part of the chromosphere in NLTE conditions.
More realistic calculations show that Mg~{\sc ii} is the dominant ionization state 
in most of the chromosphere and that, as a consequence, there is a residual but important increase
in the column density of Mg~{\sc i}. 
Correcting for the amount of Mg~{\sc i} in the chromosphere slightly raises the height of formation at disk center 
of the $b$-lines, but has an important impact for oblique observations close to the limb.
The lines then form at $\sim $800-1000~km, a region where the collisional rates are slightly
larger than the radiative rates of the metastable lower levels of this triplet $\tau^{-1}_{\rm life}({}^3P^\circ)$, yet
not large enough to completely depolarize them. 
This is important because the Mg~{\sc i} $b$-lines show a clear scattering 
polarization pattern \citep{Stenflo+00} whose formation, has been argued, could
only be understood by the presence of atomic polarization in the metastable ${}^3P^\circ$ levels
 \citep{Trujillo01}.
}

The Sr~{\sc i} resonant line at 460.7~nm has been extensively used
for the diagnostics of unresolved/disorganized/turbulent fields in
the solar atmosphere through the Hanle effect
 \citep[e.g.,][]{Stenflo82, Faurobert93, Faurobert+01, Trujillo+04, Bommier+05}.
One of the reasons for this interest is that it shows one of the 
largest linear polarization signals in the visible solar limb.
The 460.7~nm line forms in the upper photosphere and the close to the limb
observations correspond to $\sim 400$~km \citep{Trujillo+04}. 
At such height, $g^{K} < \tau_{\rm life}^{-1}$ (Figure~\ref{sunny})
and depolarizing collisions cannot destroy the atomic polarization generated
in the strong resonance line. Yet, they are of the same order of magnitude
and collisions modulate the observed signal. 
Therefore, accurate depolarizing rates are necessary for precise measurements 
of the turbulent magnetic fields in the solar photosphere.


\section{Conclusions}

The main results of this work are summarized in Table~\ref{depolarization-constants} and Figure~\ref{sunny}.
Table~\ref{depolarization-constants} gives the effective depolarizing collisional rates averaged over 
a Maxwellian distribution of velocities for the colliders ($T\leq 10000$~K),
for low lying levels of four neutral and singly ionized alkalines of 
astrophysical relevance. The $C^{(K)}(\alpha J'\leftarrow \alpha J )$ collisional rates needed
to solve the master equations, Eq.(\ref{master-equation-densities}), with 
the radiative terms required for a complete treatment, have been fitted and the corresponding parameters
are listed in the appendix.

The cross-sections have been computed from interatomic potentials calculated using
the most up-to-date {\it ab initio} methods using an adiabatic approach, and the rates
using a quantum time-independent close coupling approach. For the excited electronic
states there are many curve crossing with ionic states which introduce complicated
features in the energy curves considered. At these crossings there are non-adiabatic
couplings which may induce inelastic transitions among different $L,S_A$ manifolds. 
Because of the many crossings observed in this work, it is concluded that it is necessary to
go beyond the adiabatic quantum {\it ab initio} method used here or the diabatic
semiclassical method \citep{Brueckner:71,Anstee-Omara:95,Derouich+03} in order to
incorporate inelastic transitions for these kind of systems, in order to
get more realistic results.

Figure~\ref{sunny} summarizes the relative importance of the 
depolarizing collisions and the (polarizing) radiative rates in 
the solar atmosphere. 
We have considered the effect of depolarizing collisions on the polarization 
pattern of resonance lines of the studied species.

\section{Acknowledgments} 
{\bf Financial support by the Spanish Ministry of Economy and Competitiveness through projects  CONSOLIDER INGENIO CSD2009-00038 (Molecular Astrophysics: The Herschel and Alma Era), \mbox{FIS2011-29596-C02} and \mbox{AYA2010--18029} is gratefully acknowledged. The access to 
the CESGA computing center, through ICTS grants, is also acknowledged. }

\bibliographystyle{apj} 

\newpage
\section{Appendix}

{\bf The modeling of the observed spectral line polarization
requires the knowledge of all the $C^{(K)}( \alpha' J'\leftarrow \alpha J)$
collisional rates.
In this work we consider the quasi-elastic (or weakly inelastic)
rates, with $\alpha'=\alpha$.
We fitted all such collisional rates by using the following analytical
function:
\begin{equation*}\label{fit}
C^{(K)}(\alpha J'\leftarrow \alpha J )=\sqrt{{ 2J'+1\over 2J+1}}\quad  a \times
10^{-9}\left(\frac{T}{5000}\right)^{b} c^{T/5000} N_{H}.
\end{equation*}
The term $c^{T/5000}$ has been added to fit some of the terms
presenting very different
slopes at low and high temperatures. In general, all the
$C^{(K)}( \alpha J\leftarrow \alpha J)$
and $C^{(0)}( \alpha J' \leftarrow \alpha J)$ (with $J'\neq J$) show
a monotonously increasing behavior which is very nicely fitted by the
functional form given above.
For $K >0$, and $J\neq J'$, there are some cases for which
the calculated $C^{(K)}( \alpha J'\leftarrow \alpha J)$ show an oscillation at
low temperatures.
Such behavior is not well reproduced without the term $c^{T/5000}$,
but in all the cases the fit is good for $T>5000$.
The fitting parameters thus obtained
for all the systems and electronic terms ($\alpha$) are listed in the
following Tables.
From these $C^{(K)}(\alpha J'\leftarrow \alpha J )$  coefficients the total
rate $\bar{C}(\alpha J)$, given in Eq.~(\ref{inelastic-rate-moments}),
and the elastic depolarization rates
$D^{(K)}(\alpha J)$, given in Eq.~(\ref{elastic-depolarization-rates}),
are easily obtained.
It is worth mentioning that the fits in Table \ref{depolarization-constants}
were obtained independently to those listed below. 
It is also important to note that 
since the fits to $C^{(K)}(\alpha J'\leftarrow \alpha J )$ and $C^{(K)}(\alpha J \leftarrow \alpha J' )$ 
were obtained independently, they do not strictly satisfy the detailed balance relationships.
}

\begin{table}[h]
\caption{\label{CK-rates-MgII} Parameters obtained for Mg II}
\begin{center}
\begin{tabular}{ccccccc}
\multicolumn{6}{c}{Mg II} \\
\hline\hline
term($\alpha$)  & $J$& $J'$ & $K$& {$a$}  &  $b$  & $c$\\
\hline
$3s, ^2S$   & 1/2 & 1/2 & 0 & 22.2697 &  0.2688 &  0.8618 \\
           & 1/2 & 1/2 & 1 & 19.8177 &  0.2720 &  0.8197 \\
\hline
$3p, ^2P$   & 1/2 & 1/2 & 0 & 16.0459 &  0.1339 &  1.0269 \\
           & 1/2 & 1/2 & 1 & 14.9229 &  0.1484 &  0.9955 \\
           & 1/2 & 3/2 & 0 &  3.3010 &  0.4964 &  0.9016 \\
           & 1/2 & 3/2 & 1 & -0.6678 &  0.6181 &  0.7443 \\
           & 3/2 & 1/2 & 0 &  2.9787 &  0.3331 &  1.0189 \\
           & 3/2 & 1/2 & 1 & -0.6683 &  0.5480 &  0.7675 \\
           & 3/2 & 3/2 & 0 & 17.6662 &  0.1534 &  1.0285 \\
           & 3/2 & 3/2 & 1 & 14.8318 &  0.1342 &  1.0157 \\
           & 3/2 & 3/2 & 2 & 14.0900 &  0.1229 &  1.0080 \\
           & 3/2 & 3/2 & 3 & 14.4004 &  0.1279 &  1.0046 \\
\hline
$4s, ^2S$  & 1/2 & 1/2 & 0 & 29.4615 &  0.2790 &  1.1046 \\
          & 1/2 & 1/2 & 1 & 22.0850 &  0.2534 &  1.1455 \\
\hline
$3d, ^2D$  & 3/2 & 3/2 & 0 & 36.5529 &  0.3559 &  0.9904 \\
          & 3/2 & 3/2 & 1 & 25.9186 &  0.3367 &  1.0004 \\
          & 3/2 & 3/2 & 2 & 21.4675 &  0.3197 &  1.0044 \\
          & 3/2 & 3/2 & 3 & 22.1888 &  0.3220 &  1.0069 \\
          & 3/2 & 5/2 & 0 & 17.5339 &  0.4005 &  0.9449 \\
          & 3/2 & 5/2 & 1 & -2.4611 &  0.3683 &  0.8772 \\
          & 3/2 & 5/2 & 2 &  2.6412 &  0.4185 &  0.9148 \\
          & 3/2 & 5/2 & 3 & -0.5544 &  0.4104 &  0.9631 \\
          & 5/2 & 3/2 & 0 & 17.5517 &  0.4021 &  0.9438 \\
          & 5/2 & 3/2 & 1 & -2.4640 &  0.3700 &  0.8761 \\
          & 5/2 & 3/2 & 2 &  2.6437 &  0.4200 &  0.9138 \\
          & 5/2 & 3/2 & 3 & -0.5549 &  0.4120 &  0.9620 \\
          & 5/2 & 5/2 & 0 & 43.7542 &  0.3627 &  0.9830 \\
          & 5/2 & 5/2 & 1 & 30.3133 &  0.3465 &  0.9971 \\
          & 5/2 & 5/2 & 2 & 24.5383 &  0.3356 &  1.0018 \\
          & 5/2 & 5/2 & 3 & 22.6828 &  0.3278 &  0.9989 \\
          & 5/2 & 5/2 & 4 & 20.2225 &  0.3160 &  1.0069 \\
          & 5/2 & 5/2 & 5 & 22.4469 &  0.3249 &  1.0059 \\
\hline\hline
\end{tabular}
\end{center}
\end{table}

\begin{table}[h]
\caption{\label{CK-rates-CaII} Parameters obtained for Ca II}
\begin{center}
\begin{tabular}{ccccccc}
\multicolumn{6}{c}{Ca II} \\
\hline\hline
term($\alpha$)  & $J$& $J'$ & $K$& {$a$}  &  $b$  & $c$\\
\hline
$4s, ^2S$  & 1/2 & 1/2 & 0 & 10.1683 &  0.1461 &  1.1290 \\
          & 1/2 & 1/2 & 1 &  6.8879 &  0.0766 &  1.1440 \\
\hline
$3d, ^2D$  & 3/2 & 3/2 & 0 & 11.6422 &  0.2537 &  0.9997 \\
          & 3/2 & 3/2 & 1 & 10.2758 &  0.2673 &  0.9805 \\
          & 3/2 & 3/2 & 2 &  9.8214 &  0.2698 &  0.9775 \\
          & 3/2 & 3/2 & 3 &  9.9750 &  0.2648 &  0.9819 \\
          & 3/2 & 5/2 & 0 &  1.8837 &  0.3232 &  1.0090 \\
          & 3/2 & 5/2 & 1 & -0.1043 &  0.2138 &  0.6809 \\
          & 3/2 & 5/2 & 2 &  0.1163 &  0.3383 &  1.2833 \\
          & 3/2 & 5/2 & 3 & -0.0608 &  0.0704 &  1.1641 \\
          & 5/2 & 3/2 & 0 &  1.7165 &  0.1947 &  1.1179 \\
          & 5/2 & 3/2 & 1 & -0.1054 &  0.1604 &  0.6896 \\
          & 5/2 & 3/2 & 2 &  0.1052 &  0.1960 &  1.4333 \\
          & 5/2 & 3/2 & 3 & -0.0543 & -0.0726 &  1.3118 \\
          & 5/2 & 5/2 & 0 & 12.2968 &  0.2641 &  0.9974 \\
          & 5/2 & 5/2 & 1 & 10.5245 &  0.2613 &  0.9883 \\
          & 5/2 & 5/2 & 2 &  9.9563 &  0.2636 &  0.9896 \\
          & 5/2 & 5/2 & 3 &  9.5741 &  0.2629 &  0.9872 \\
          & 5/2 & 5/2 & 4 &  9.4489 &  0.2622 &  0.9842 \\
          & 5/2 & 5/2 & 5 &  9.7655 &  0.2550 &  0.9933 \\
\hline
$4p, ^2P$  & 1/2 & 1/2 & 0 & 10.6365 &  0.1941 &  1.0711 \\
          & 1/2 & 1/2 & 1 &  8.5475 &  0.2386 &  1.0475 \\
          & 1/2 & 3/2 & 0 &  3.8173 &  0.7415 &  0.8225 \\
          & 1/2 & 3/2 & 1 & -0.8538 &  1.2265 &  0.5517 \\
          & 3/2 & 1/2 & 0 &  3.3951 &  0.4626 &  0.9852 \\
          & 3/2 & 1/2 & 1 & -0.8643 &  1.0820 &  0.5869 \\
          & 3/2 & 3/2 & 0 & 13.7498 &  0.3205 &  1.0098 \\
          & 3/2 & 3/2 & 1 &  9.5825 &  0.3292 &  1.0144 \\
          & 3/2 & 3/2 & 2 &  8.4721 &  0.3098 &  1.0102 \\
          & 3/2 & 3/2 & 3 &  9.3266 &  0.3349 &  0.9747 \\
\hline\hline
\end{tabular}
\end{center}
\end{table} 

\begin{table}[h]
\caption{\label{CK-rates-BaII}
Parameters obtained for Ba II, for the barium isotopes with zero nuclear spin,  $^{138}$Ba  and $^{136}$Ba}
\begin{center}
\begin{tabular}{ccccccc}
\multicolumn{6}{c}{Ba II} \\
\hline\hline
term($\alpha$)  & $J$& $J'$ & $K$& {$a$}  &  $b$  & $c$\\
\hline
$6s, ^2S$  & 1/2 & 1/2 &  0 & 13.9609 &  0.1386 &  1.0841 \\
          & 1/2 & 1/2 &  1 &  9.9264 &  0.0790 &  1.0680 \\
\hline
$5d, ^2D$  & 3/2 & 3/2 &  0 & 18.4255 &  0.2735 &  0.9255 \\
          & 3/2 & 3/2 &  1 & 15.7971 &  0.3207 &  0.8943 \\
          & 3/2 & 3/2 &  2 & 14.7100 &  0.3087 &  0.9012 \\
          & 3/2 & 3/2 &  3 & 14.3710 &  0.3080 &  0.9085 \\
          & 3/2 & 5/2 &  0 &  1.7446 &  1.4666 &  0.6342 \\
          & 3/2 & 5/2 &  1 & -0.2700 &  2.1136 &  0.4071 \\
          & 3/2 & 5/2 &  2 &  0.1545 &  2.3241 &  0.4218 \\
          & 3/2 & 5/2 &  3 &  0.0907 &  3.3058 &  0.1347 \\
          & 5/2 & 3/2 &  0 &  1.6072 &  0.8165 &  0.8888 \\
          & 5/2 & 3/2 &  1 & -0.2804 &  1.6079 &  0.5086 \\
          & 5/2 & 3/2 &  2 &  0.1575 &  1.7971 &  0.5363 \\
          & 5/2 & 3/2 &  3 &  0.1246 &  3.1623 &  0.1275 \\
          & 5/2 & 5/2 &  0 & 17.1487 &  0.2515 &  0.9693 \\
          & 5/2 & 5/2 &  1 & 13.9871 &  0.2599 &  0.9609 \\
          & 5/2 & 5/2 &  2 & 13.4622 &  0.2621 &  0.9545 \\
          & 5/2 & 5/2 &  3 & 12.7033 &  0.2530 &  0.9578 \\
          & 5/2 & 5/2 &  4 & 12.3177 &  0.2512 &  0.9629 \\
          & 5/2 & 5/2 &  5 & 13.2108 &  0.2582 &  0.9556 \\
\hline
$6p, ^2P$  & 1/2 & 1/2 &  0 & 17.8692 &  0.2161 &  1.0473 \\
          & 1/2 & 1/2 &  1 &  8.8049 &  0.1112 &  1.0957 \\
          & 1/2 & 3/2 &  0 &  0.5194 &  1.5731 &  0.7797 \\
          & 1/2 & 3/2 &  1 & -0.0132 &  1.9483 &  0.9342 \\
          & 3/2 & 1/2 &  0 &  0.4897 &  0.3255 &  1.4292 \\
          & 3/2 & 1/2 &  1 & -0.0118 &  0.6236 &  1.8070 \\
          & 3/2 & 3/2 &  0 & 19.7847 &  0.2930 &  1.0269 \\
          & 3/2 & 3/2 &  1 & 11.8792 &  0.2537 &  1.0423 \\
          & 3/2 & 3/2 &  2 & 10.3170 &  0.2242 &  1.0267 \\
          & 3/2 & 3/2 &  3 & 10.9570 &  0.2434 &  1.0275 \\
\hline\hline
\end{tabular}
\end{center}
\end{table} 

\begin{table}[h]
\caption{\label{CK-rates-MgI}
Parameters obtained for Mg I }
\begin{center}
\begin{tabular}{ccccccc}
\multicolumn{6}{c}{Mg I} \\
\hline\hline
term($\alpha$)  & $J$& $J'$ & $K$& {$a$}  &  $b$  & $c$\\
\hline
$3s3p, ^3P$  & 0 & 0 &  0 &  5.2652 &  0.3164 &  1.0542 \\
            & 0 & 1 &  0 &  1.4108 &  0.3435 &  1.0455 \\
            & 0 & 2 &  0 &  1.9195 &  0.5142 &  0.9203 \\
            & 1 & 0 &  0 &  1.3689 &  0.3011 &  1.0811 \\
            & 1 & 1 &  0 &  6.7890 &  0.3075 &  1.0614 \\
            & 1 & 1 &  1 &  5.1094 &  0.3035 &  1.0548 \\
            & 1 & 1 &  2 &  4.7392 &  0.2933 &  1.0632 \\
            & 1 & 2 &  0 &  3.2952 &  0.4699 &  0.9504 \\
            & 1 & 2 &  1 &  1.1216 &  0.5025 &  0.9375 \\
            & 1 & 2 &  2 & -0.4384 &  0.4947 &  0.8861 \\
            & 2 & 0 &  0 &  1.8167 &  0.4168 &  0.9868 \\
            & 2 & 1 &  0 &  3.1625 &  0.4009 &  0.9996 \\
            & 2 & 1 &  1 &  1.0850 &  0.4405 &  0.9792 \\
            & 2 & 1 &  2 & -0.4205 &  0.4256 &  0.9323 \\
            & 2 & 2 &  0 &  8.5357 &  0.3056 &  1.0543 \\
            & 2 & 2 &  1 &  6.9518 &  0.2988 &  1.0563 \\
            & 2 & 2 &  2 &  5.1478 &  0.2619 &  1.0745 \\
            & 2 & 2 &  3 &  4.7209 &  0.2446 &  1.0760 \\
            & 2 & 2 &  4 &  4.5807 &  0.2306 &  1.0815 \\
\hline
$3s3p, ^1P$  & 1 & 1 &  0 & 21.8188 &  0.3833 &  1.0077 \\
            & 1 & 1 &  1 &  8.1003 &  0.3276 &  1.0240 \\
            & 1 & 1 &  2 & 10.1407 &  0.3450 &  0.9925 \\
\hline
$3s4s, ^3S$  & 1 & 1 &  0 & 39.5285 &  0.4225 &  0.9836 \\
            & 1 & 1 &  1 & 33.0867 &  0.4151 &  0.9875 \\
            & 1 & 1 &  2 & 20.2220 &  0.3876 &  1.0022 \\
\hline\hline
\end{tabular}
\end{center}
\end{table}

\begin{table}[h]
\caption{\label{CK-rates-SrI}
Parameters obtained for Sr I }
\begin{center}
\begin{tabular}{ccccccc}
\multicolumn{6}{c}{Sr I} \\
\hline\hline
term($\alpha$)  & $J$& $J'$ & $K$& {$a$}  &  $b$  & $c$\\
\hline
$5s5p, ^1P$  & 1 & 1 &  0 & 43.4158 &  0.4514 &  0.8079 \\
             & 1 & 1 &  1 & 34.6220 &  0.5641 &  0.7091 \\
             & 1 & 1 &  2 & 37.0942 &  0.5247 &  0.7198 \\
\hline\hline
\end{tabular}
\end{center}
\end{table}

\begin{table}[h]
\caption{\label{CK-rates-BaI}
Parameters obtained for Ba I, 
for the barium isotopes with zero nuclear spin,  $^{138}$Ba  and $^{136}$Ba}
\begin{center}
\begin{tabular}{ccccccc}
\multicolumn{6}{c}{Ba I} \\
\hline\hline
term($\alpha$)  & $J$& $J'$ & $K$& {$a$}  &  $b$  & $c$\\
\hline
$6s5d, ^3P$  & 1 & 1 &  0 & 14.3162 &  0.2865 &  1.0169 \\
            & 1 & 1 &  1 & 11.6651 &  0.3703 &  0.9979 \\
            & 1 & 1 &  2 &  9.6991 &  0.4113 &  0.9922 \\
            & 1 & 2 &  0 &  3.8751 &  0.7136 &  0.7996 \\
            & 1 & 2 &  1 &  1.0372 &  1.2164 &  0.7024 \\
            & 1 & 2 &  2 &  0.6504 &  0.9258 &  0.7962 \\
            & 1 & 3 &  0 &  1.7184 &  1.2332 &  0.6865 \\
            & 1 & 3 &  1 & -0.4099 &  1.6277 &  0.6003 \\
            & 1 & 3 &  2 &  0.0648 &  1.5280 &  0.6417 \\
            & 2 & 1 &  0 &  3.5544 &  0.4957 &  0.9187 \\
            & 2 & 1 &  1 &  1.0060 &  1.0547 &  0.7669 \\
            & 2 & 1 &  2 &  0.6113 &  0.7308 &  0.8949 \\
            & 2 & 2 &  0 & 12.5009 &  0.2241 &  1.0888 \\
            & 2 & 2 &  1 &  9.2796 &  0.2221 &  1.1302 \\
            & 2 & 2 &  2 &  9.0408 &  0.2330 &  1.1230 \\
            & 2 & 2 &  3 &  8.5488 &  0.2293 &  1.1303 \\
            & 2 & 2 &  4 &  7.7708 &  0.2212 &  1.1431 \\
            & 2 & 3 &  0 &  3.6769 &  1.0237 &  0.7401 \\
            & 2 & 3 &  1 &  1.5593 &  1.3927 &  0.6463 \\
            & 2 & 3 &  2 &  0.8961 &  1.3228 &  0.6951 \\
            & 2 & 3 &  3 &  0.7816 &  1.3829 &  0.6559 \\
            & 2 & 3 &  4 &  0.3919 &  1.1992 &  0.7131 \\
            & 3 & 1 &  0 &  1.6158 &  0.7694 &  0.8741 \\
            & 3 & 1 &  1 & -0.4331 &  1.3037 &  0.6834 \\
            & 3 & 1 &  2 &  0.0534 &  0.9093 &  0.9270 \\
            & 3 & 2 &  0 &  3.2868 &  0.6344 &  0.9303 \\
            & 3 & 2 &  1 &  1.5085 &  1.0888 &  0.7550 \\
            & 3 & 2 &  2 &  0.8403 &  0.9830 &  0.8362 \\
            & 3 & 2 &  3 &  0.7589 &  1.0831 &  0.7636 \\
            & 3 & 2 &  4 &  0.3883 &  0.9204 &  0.8149 \\
            & 3 & 3 &  0 & 16.3603 &  0.3644 &  0.9208 \\
            & 3 & 3 &  1 & 14.6417 &  0.3896 &  0.9057 \\
            & 3 & 3 &  2 & 13.1621 &  0.3815 &  0.9045 \\
            & 3 & 3 &  3 & 11.8795 &  0.3712 &  0.9119 \\
            & 3 & 3 &  4 & 10.5420 &  0.3580 &  0.9228 \\
            & 3 & 3 &  5 &  9.0805 &  0.3424 &  0.9458 \\
            & 3 & 3 &  6 &  7.3338 &  0.3232 &  0.9733 \\
\hline
$6s5d, ^1D$  & 2 & 2 &  0 & 24.0293 &  0.4057 &  0.9585 \\
            & 2 & 2 &  1 & 15.1879 &  0.4196 &  0.8876 \\
            & 2 & 2 &  2 & 14.4604 &  0.4057 &  0.8988 \\
            & 2 & 2 &  3 & 13.2916 &  0.4144 &  0.8892 \\
            & 2 & 2 &  4 & 14.5496 &  0.4049 &  0.9003 \\
\hline
$6s6p, ^3P$  & 0 & 0 &  0 & 36.4898 &  0.5360 &  0.7526 \\
            & 0 & 1 &  0 &  4.2507 &  1.0777 &  0.7470 \\
            & 0 & 2 &  0 &  0.6928 &  2.0524 &  0.5844 \\
            & 1 & 0 &  0 &  3.9279 &  0.7297 &  0.9079 \\
            & 1 & 1 &  0 & 20.0150 &  0.2993 &  0.8131 \\
            & 1 & 1 &  1 & 12.0838 &  0.3141 &  0.8287 \\
            & 1 & 1 &  2 &  8.4330 &  0.2368 &  0.8777 \\
            & 1 & 2 &  0 &  1.6280 &  1.7001 &  0.6576 \\
            & 1 & 2 &  1 &  0.2266 &  1.3048 &  0.8525 \\
            & 1 & 2 &  2 & -0.6519 &  4.0471 &  0.0337 \\
            & 2 & 0 &  0 &  0.7328 &  1.2606 &  0.8290 \\
            & 2 & 1 &  0 &  1.6054 &  1.0751 &  0.8861 \\
            & 2 & 1 &  1 &  0.2249 &  0.6818 &  1.1432 \\
            & 2 & 1 &  2 & -0.5533 &  3.2715 &  0.0523 \\
            & 2 & 2 &  0 & 24.1559 &  0.4468 &  0.9130 \\
            & 2 & 2 &  1 & 17.4109 &  0.4954 &  0.9064 \\
            & 2 & 2 &  2 & 13.2654 &  0.5063 &  0.8978 \\
            & 2 & 2 &  3 & 12.0040 &  0.4861 &  0.9038 \\
            & 2 & 2 &  4 & 11.1925 &  0.4619 &  0.8975 \\
\hline
$6s6p, ^1P$  & 1 & 1 &  0 & 31.4136 &  0.3874 &  1.0268 \\
            & 1 & 1 &  1 & 12.9017 &  0.3391 &  1.0644 \\
            & 1 & 1 &  2 & 15.1697 &  0.3536 &  1.0498 \\
\hline\hline
\end{tabular}
\end{center}
\end{table} 

\end{document}